\documentclass[superscriptaddress,showpacs,twocolumn,amsmath,amssymb,pra,longbibliography]{revtex4-1}

\usepackage{graphicx}
\usepackage{dcolumn}
\usepackage{bm}
\usepackage{color}
\usepackage[unicode]{hyperref}
\hypersetup{
   unicode=true,          
   plainpages=false,
   colorlinks=true,       
   citecolor=blue,        
}
\newcommand{\rref}[1]{Eq.\ (\ref{#1})}
\newcommand{\rrefsa}[1]{Eqs.\ (\ref{#1})}
\newcommand{\rrefsb}[1]{(\ref{#1})}
\newcommand{\erf}{\mathop{\mathrm{erf}}}

\begin{document}

\preprint{AIP/123-QED}

\title[The $s$-wave scattering length of a Gaussian potential]{The $s$-wave scattering length of a Gaussian potential}

\author{Peter Jeszenszki}
 \affiliation{Dodd-Walls Centre for Photonics and Quantum Technology,
New Zealand Institute for Advanced Study, and Centre for Theoretical Chemistry and Physics, Massey University, Private Bag 102904 North Shore,  Auckland 0745, New Zealand}
\author{Alexander Yu.\ Cherny}%
\affiliation{ Bogoliubov Laboratory of Theoretical Physics, Joint Institute for Nuclear Research, 141980 Dubna, Russia 
}%

\author{Joachim Brand}
\email{J.Brand@massey.ac.nz}
\affiliation{Dodd-Walls Centre for Photonics and Quantum Technology,
New Zealand Institute for Advanced Study, and Centre for Theoretical Chemistry and Physics, Massey University, Private Bag 102904 North Shore,  Auckland 0745, New Zealand}
\affiliation{Max Planck Institute for Solid State Research, Heisenbergstra{\ss}e 1, 70569 Stuttgart, Germany}


\date{\today}
\begin{abstract}
We provide accurate expressions for the $s$-wave scattering length for a Gaussian potential well in one, two and three spatial dimensions. The Gaussian potential is widely used
as a pseudopotential
in the theoretical description of ultracold atomic gases, where the $s$-wave scattering length is a physically relevant parameter. 
We first describe a numerical procedure to compute the value of the $s$-wave scattering length from the parameters of the Gaussian but find that its accuracy is limited in the vicinity of singularities that result from the formation of new bound states.
We then derive simple analytical expressions that capture the correct asymptotic behavior of the $s$-wave scattering length near the bound states. Expressions that are increasingly accurate in wide parameter regimes are found by a hierarchy of approximations that capture an increasing number of bound states. The small number of numerical coefficients that enter these expressions is determined from accurate numerical calculations.
The approximate formulas combine the advantages of the numerical and approximate 
expressions, yielding an accurate and simple description from the weakly to the strongly 
interacting limit.
\end{abstract}

\maketitle

\section{\label{sec:intro}Introduction}

The interest in the accurate determination of $s$-wave scattering length
has increased in recent decades due to  its importance in the  
description of systems of ultracold atoms \cite{pethick_bose-einstein_2002,pitaevskii16:book}. 
As the range of the interparticle interactions is usually much smaller than the average 
inter-particle distances, the effects of interactions can be expressed in terms of the scattering amplitude between pairs of particles. 
For dilute gases at ultracold temperatures, the kinetic energies are low and, therefore, the main contribution to the amplitude comes from the $s$-wave scattering at 
zero momentum. Particle interactions are thus  determined completely by a single parameter: the $s$-wave scattering 
length \cite{roger_g._newton_scattering_1982,pethick_bose-einstein_2002}. In theoretical calculations, it is therefore not necessary to consider the detailed interaction potential between 
the particles. Instead, a pseudopotential may be chosen in a way to reproduce the 
desired value of the $s$-wave scattering length, which can simplify the required computations considerably \cite{pethick_bose-einstein_2002}.

One of the simplest and most popular pseudopotentials is the Dirac $\delta$ potential. Its straightforward application is, however,
restricted to one dimension, since in two or three dimensions it is meaningless without renormalization
\cite{esry_validity_1998,rontani_configuration_2013,doganov_two_2013}.
An alternative option is to use finite-range pseudopotentials, 
e.g.\ the finite square well~\cite{stecher_energetics_2008,blume_few-body_2012}, 
Troullier-Martins~\cite{bugnion_high-fidelity_2014,whitehead_pseudopotential_2016}, 
P\"oschl-Teller~\cite{forbes_resonantly_2011,galea_diffusion_2016}, or
Gaussian potential \cite{stecher_energetics_2008,blume_few-body_2012,doganov_two_2013,christensson_effective-interaction_2009,
klaiman_breaking_2014,beinke_many-body_2015,imran_exact_2015,Bolsinger_2016,Bolsinger_2017}. The scattering length is finite for these
pseudopotentials, but an extrapolation to zero range might be necessary to avoid an unphysical 
shape dependence~\cite{stecher_energetics_2008,blume_few-body_2012,forbes_resonantly_2011}.  
The relationship between the parameter(s) of the potential and 
the scattering length is not always trivial. Apart from some 
special cases~\cite{farrell_s-wave_2010,forbes_resonantly_2011}, 
numerical techniques are required to determine this relation~\cite{landau_course_1977,verhaar_scattering_1985,Galea_2017}.


For Gaussian potentials, no closed-form analytic expressions are available and,
 for this reason, numerical approaches have been applied 
~\cite{christensson_effective-interaction_2009,parish_bcs-bec_2005,johnson_effective_2012,Bolsinger_2016}. In two dimensions, an
approximate expression was derived by Doganov \emph{et al.}~\cite{doganov_two_2013}.
These authors considered two particles in a harmonic trap, where the Gaussian interparticle
interaction is treated in a perturbative framework. The obtained second order correction
combined with the analytical result of the contact pseudopotential
~\cite{busch_two_1998,farrell_universality_2010} provides the approximate
expression. Due to the perturbative approach, this approximation works quite well in the 
weakly interacting limit, but it deteriorates with increasing  interaction strength. 

In this paper, we derive  approximate analytical expressions for the $s$-wave scattering length 
of a Gaussian pseudopotential in one, two
and three dimensions. 
These expressions qualitatively describe the singularities of
the $s$-wave scattering length at the formation of the first bound state, which is problematic for 
purely numerical approaches. Analytical formulas for weak interactions are derived in one and two dimensions, where the $s$-wave scattering length has a singularity
at zero interaction strength.
In order to improve the accuracy, the approximate expressions are generalized
by including the effects of additional bound states. The unknown parameters
in this ansatz are determined by non-linear fitting to accurate numerical results.
The obtained formulas are robust and simple and accurately provide the values for the $s$-wave scattering length in a wide regime of attractive interaction.

We describe and carefully benchmark a numerical method to accurately determine the scattering length of a short-range scattering potential in one, two, and three spatial dimensions. The approach is based on previous work of Verhaar \cite{verhaar_scattering_1985} and may be useful in its own right as it is able to provide very accurate results except for the immediate vicinity of the singularities. The numerical approach is applicable for general short-range potentials and is not restricted to potentials of Gaussian shape. 

This paper is organized as follows: In Sec.\ \ref{sec:Numerical}, after stating the problem and discussing the required asymptotic conditions for scattering wave functions, we present an accurate numerical approach for determining the $s$-wave scattering length along with benchmark calculations for a Gaussian potential.
In Sec.\ \ref{sec:Approx} approximate analytic expressions for the $s$-wave scattering length of a Gaussian potential are derived before more accurate, generalized expressions with numerically determined parameters are introduced.  Three appendices provide additional details on derivations and numerical issues with determining the position of singularities in the scattering length, respectively.




\section{\label{sec:Numerical}Numerical determination of the $S$-wave scattering length}

\subsection{Solution of the two-body problem and connection with the $s$-wave scattering length}
\subsubsection{Two-body scattering problem}\label{sec:2bp}
Let us consider a two-particle scattering process with the following $n$-dimensional  Hamiltonian:
\begin{eqnarray}
H_{2p} &=& - \frac{\hbar^2}{2m_1}\nabla^2_1  - \frac{\hbar^2}{2m_2} \nabla^2_2 + V( | {\mathbf r}_1 - {\mathbf r}_2 |) \ ,
\label{2part}
\end{eqnarray}
where $V( | {\mathbf r}_1 - {\mathbf r}_2 |)$ is a spherically symmetric particle-particle interaction potential, and
$m_i$, ${\mathbf r}_i$, and $\nabla^2_i$ are the mass,  coordinate, and  
Laplace operator
of the $i$th particle, respectively. Although our main target is the Gaussian potential, here we consider more general classes of potentials for which the numerical procedures can be applied.
Specifically, we assume that the interaction is sufficiently short ranged to justify the existence of the scattering length.
This is fulfilled in $n$ dimensions if $V(r)$ obeys the condition~\cite{case50,frank71} 
\begin{align*}
\int\limits_A^{\infty} |V(r)|r^{n-1} \mbox{d} r<\infty
\end{align*} 
for a finite $A$.
It is sufficient to assume that the potential decreases faster than $1/r^{n+\varepsilon}$ with $\varepsilon>0$ at sufficiently large distance.
In addition, we suppose that the potential is regular at the origin or 
diverges, at most, with $1/r^s$ with $s<1$.
This condition is necessary to uniquely define the appropriate boundary conditions at the origin for the purpose of the numerical procedure. The $s$-wave scattering length can still be defined for more strongly divergent potentials \cite{frank71,andrews76},
 but the numerical procedure would have to be modified in this case.

The eigenproblem for the Hamiltonian (\ref{2part}) can be simplified by 
introducing the center of mass coordinate, 
${\mathbf R}=(m_1{\mathbf r}_1+m_2{\mathbf r}_2)/(m_1+m_2)$,  and relative coordinate, ${\mathbf r}={\mathbf r}_1-{\mathbf r}_2$. Then the wave function can be separated as ~\cite{roger_g._newton_scattering_1982}
\begin{equation}
\Psi_{nD}^{2p}({\bf r}_1, {\bf r}_2) = \exp(i{\bf Q}\cdot{\bf R}/\hbar) \, \psi_{nD}( {\bf r}) \ , \nonumber
\end{equation}
where ${\bf Q}$ is the total momentum of the two particles. The relative wave function  $\psi_{nD}( {\bf r})$  is an eigenfunction of the Hamiltonian of the relative
motion,
\begin{eqnarray}
H\, \Psi_{nD}( {\bf r}) &=& E \, \Psi_{nD}( {\bf r}) \ \label{fullrelsch}
\end{eqnarray}
with
\begin{align}
H = -\frac{\hbar^2}{2 \mu} \nabla^2 + V(r) \ ,
\end{align}
where $E$ is the scattering energy and $\mu = m_1 m_2/(m_1+m_2)$ is the reduced mass.

\subsubsection{The $s$-wave scattering and boundary conditions}
Due to the spherical symmetry of the potential,  Eq.~\rrefsb{fullrelsch} can be further simplified by solving the angular-coordinate-dependent part 
separately through eigenstates of the angular momentum operator. By definition,  $s$-wave scattering correspond to zero angular momentum with a radially symmetric
wave function.
The radial coordinate dependence in  \rref{fullrelsch} 
 can be obtained from the following differential equation for the general $n$-dimensional case  ~\cite{verhaar_scattering_1985,utoPhi}:
\begin{align}
-\frac{\hbar^2}{2\mu}\left( \frac{\mbox{d}^2}{\mbox{d}r^2} + \frac{n-1}{r}
\frac{\mbox{d}}{\mbox{d}r}  \right)\Phi_{nD}( r)& + 
\label{relsch} \\
+ \left( V(r)- E \right) \Phi_{nD}&( r)=0 \ , \nonumber
\end{align}
where $\Phi_{nD}(r)$ is the radial part of the relative wave function $\psi^{rel}_{nD}( {\bf r})$. 
For ultracold atoms, only low-energy scattering processes are relevant and we may set $E=0$. As we see in the following  section, it also provides us with a 
simple way to define the $s$-wave scattering length.
Appropriate 
boundary conditions for the differential equation \rrefsb{relsch} can be obtained from smoothness and symmetry considerations in the limit 
$r \rightarrow 0$ (see the detailed description in Appendix \ref{AppendixBC}), as
\begin{eqnarray} \label{eq:bcs}
\Phi_{nD}(0)=1 \ ,    &\hspace{1cm}\mbox{and}\hspace{1cm}& \Phi_{nD}'(0)=0 \ . \label{boundarycond}
\end{eqnarray}

\subsubsection{Scattering length}

For a short-range potential, the asymptotic of the wave function $\Phi_{nD}(r)$ at  distances much larger than the characteristic length scale of the potential 
$\ell_v$ is given by a solution of Eq.~\eqref{relsch} with $V(r)=0$ and $E=0$, which is a linear combination of a constant and $r$ in one dimension (1D), $\mbox{ln} \left(r\right)$ in 2D, and $1/r$ in 3D. The $s$-wave scattering length is defined by the ratio of the corresponding constants in this linear combination \cite{verhaar_scattering_1985,utoPhi},
\begin{equation}
\mbox{if    } r \gg \ell_v 
\begin{cases}
\Phi_{1D}( r) \approx {\mathcal N}_{1D} \left( r - a^{1D}_s  \right) \ , \\
\Phi_{2D}( r) \approx {\mathcal N}_{2D}  \left[ \mbox{ln} \left( \frac{2r}{a^{2D}_s}\right)-\gamma \right]  \ ,  \\
\Phi_{3D}( r) \approx {\mathcal N}_{3D} \left( 1 - \frac{a^{3D}_s}{r}  \right) \ . 
\end{cases}
\label{assPhy}
\end{equation}
Here  $a^{nD}_s$ is the $n$-dimensional $s$-wave scattering length, 
$\gamma = 0.5772 \ldots$ is the Euler-Mascheroni constant, and $\mathcal{N}_{nD}$ is a scalar factor.
The scattering length can be expressed as a limit of the function $\Phi_{nD}(r)$ and its first derivative 
by eliminating the unknown parameter 
$\mathcal{N}_{nD}$ \cite{verhaar_scattering_1985,a2ddef,utoPhi} as
\begin{eqnarray}
a^{1D}_s &=& \lim_{r \rightarrow \infty} \left( r-\frac{\Phi_{1D}( r)}{\Phi_{1D}'( r)} \right) \ , \label{as1dnum} \\
a^{2D}_s &=& \lim_{r \rightarrow \infty} 2r \exp{\left(-\frac{\Phi_{2D}(r)}{r\Phi_{2D}'(r)}-\gamma\right)} \ , \label{as2dnum} \\
a^{3D}_s &=& \lim_{r \rightarrow \infty} \left( r-\frac{r\Phi_{3D}( r)}{r\Phi_{3D}'( r)+ \Phi_{3D}( r)} \right) \ . \label{as3dnum}
\end{eqnarray}
As can be seen from the expressions above, $a^{2D}_s$ is always positive by definition, while $a^{1D}_s$ and $a^{3D}_s$ can be of either sign.
In the limiting case where the scattering potential is absent the solution of  Eq.~\eqref{relsch} becomes a zero-energy plane wave, i.e.~the constant 1.
Therefore, we have $a^{1D}_s\to \infty$ and $a^{2D}_s\to \infty$, while $a^{3D}_s\to0$. This means that the scattering length develops a singularity when $V(r)\to0$ in one and two dimensions.

\subsection{One and three dimensions}

The radial Schr\"odinger equation can be simplified by introducing the functions~\cite{verhaar_scattering_1985}
\begin{align}
u_{3D}( r) &=  r \, \Phi_{3D}( r) \quad \textrm{and} \label{u3ddef} \\
u_{1D}(r) &=  \Phi_{1D}(r).
\end{align}
Substituting \rref{u3ddef} into the radial Schr\"odinger equation \rrefsb{relsch} and the 
expression for the $s$-wave scattering length \rrefsb{as3dnum}, we obtain identical equations for three and one dimensions,
\begin{align}
\left( -\frac{\hbar^2}{2\mu} \frac{\mbox{d}^2}{\mbox{d}r^2}
+ V(r)- E  \right) u_{1D/3D}( r) =0 \ , \label{urelsch} \\
a^{1D/3D}_s = \lim_{r \rightarrow \infty} \left( r-\frac{u_{1D/3D}( r)}{u_{1D/3D}'( r)} \right) \ . \label{localas3d}
\end{align}
The boundary conditions are 
obtained by substituting \rref{u3ddef} into \rref{boundarycond} and now differ between one and three dimensions,
\begin{eqnarray}
u_{1D}(0)=1 \ , &\hspace{2cm}& u_{1D}'(0) = 0 \ , \label{bound1d} \\
u_{3D}(0)=0 \ , &\hspace{2cm}& u_{3D}'(0) = 1 \ . \label{bound3d} 
\end{eqnarray}
 
In a numerical procedure, we may assume that the  functions $u_{1D/3D}(r)$ and $u_{1D/3D}'( r)$ can only be given with limited numerical accuracy ($p$) as
\begin{eqnarray*}
u_{1D/3D}(r)&=&\lim\limits_{p \rightarrow \infty} \tilde{u}_{1D/3D}(r;p) \ ,\\
u_{1D/3D}'(r)&=&\lim\limits_{p \rightarrow \infty} \tilde{u}_{1D/3D}'(r;p) \ ,
\end{eqnarray*}
where $p$ relates to the accuracy of the decimal representation and the numerical method itself.
For the numerical determination of the scattering length, one should then consider the combined limit,
\begin{eqnarray*}
a^{1D/3D}_s &=& \lim_{r,p \rightarrow \infty} 
\underbrace{\left( r-\frac{\tilde{u}_{1D/3D}( r;p)}{\tilde{u}_{1D/3D}'( r;p)} \right)}_{\tilde{a}_s^{1D/3D}(r;p)} \ .
\end{eqnarray*}

\subsection{Two-dimensional case}

In two dimensions, the original radial function $\Phi_{2D}(r)$ is used directly.
Here a numerical instability is present as a result of the $1/r$ singularity
in the first derivative term of the radial Schr\"odinger equation \rrefsb{relsch} for two dimensions.
The instability can be avoided by giving the boundary conditions at distance $r=\epsilon$, 
where $\epsilon$ is chosen large enough
to avoid the numerical difficulties but
small enough to approximately satisfy the conditions of Eq.~\eqref{eq:bcs}
\begin{eqnarray}
\Phi_{2D}(\epsilon) \approx 1 \ , \hspace{2cm} \Phi'_{2D}(\epsilon) \approx 0 \ . \label{initcond}
\end{eqnarray}
Consequently, $\epsilon$ becomes another parameter of the numerical evaluation besides 
the numerical accuracy ($p$). The scattering length is then obtained from the composite limit
\begin{eqnarray}
a_s^{2D} &=&
\lim_{\stackrel{r,p \rightarrow \infty}{\epsilon \rightarrow 0}} 
\underbrace{2r \exp{\left(-\frac{\tilde{\Phi}_{2D}(r;\epsilon,p)}{r\tilde{\Phi}_{2D}'(r;\epsilon,p)}-\gamma\right)}}_{\tilde{a}_s^{2D}(r;\epsilon,p)}
\label{assympexp} \ ,
\end{eqnarray}
where $\tilde{\Phi}_{2D}(r;\epsilon,p)$ represents the  approximate numerical solution of \rrefsa{relsch} and 
\rrefsb{initcond} with
\begin{eqnarray}
\Phi_{2D}(r)&=&\lim\limits_{\stackrel{p \rightarrow \infty}{\epsilon \rightarrow 0}} \tilde{\Phi}_{2D}(r;\epsilon,p) \ . \label{numr2d}
\end{eqnarray}

\subsection{The Gaussian potential and the convergence of numerical results}

\begin{figure}
\center
\includegraphics[scale=0.5]{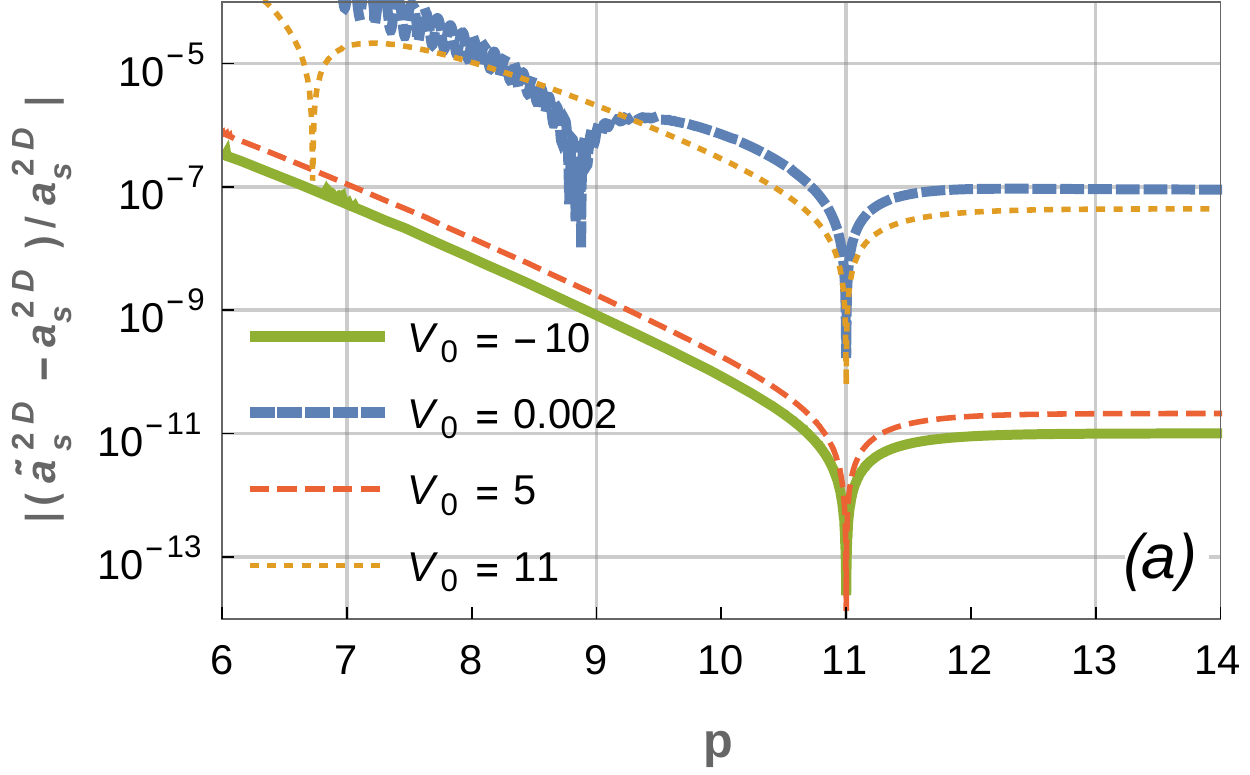} \hspace{0.5cm}
\includegraphics[scale=0.5]{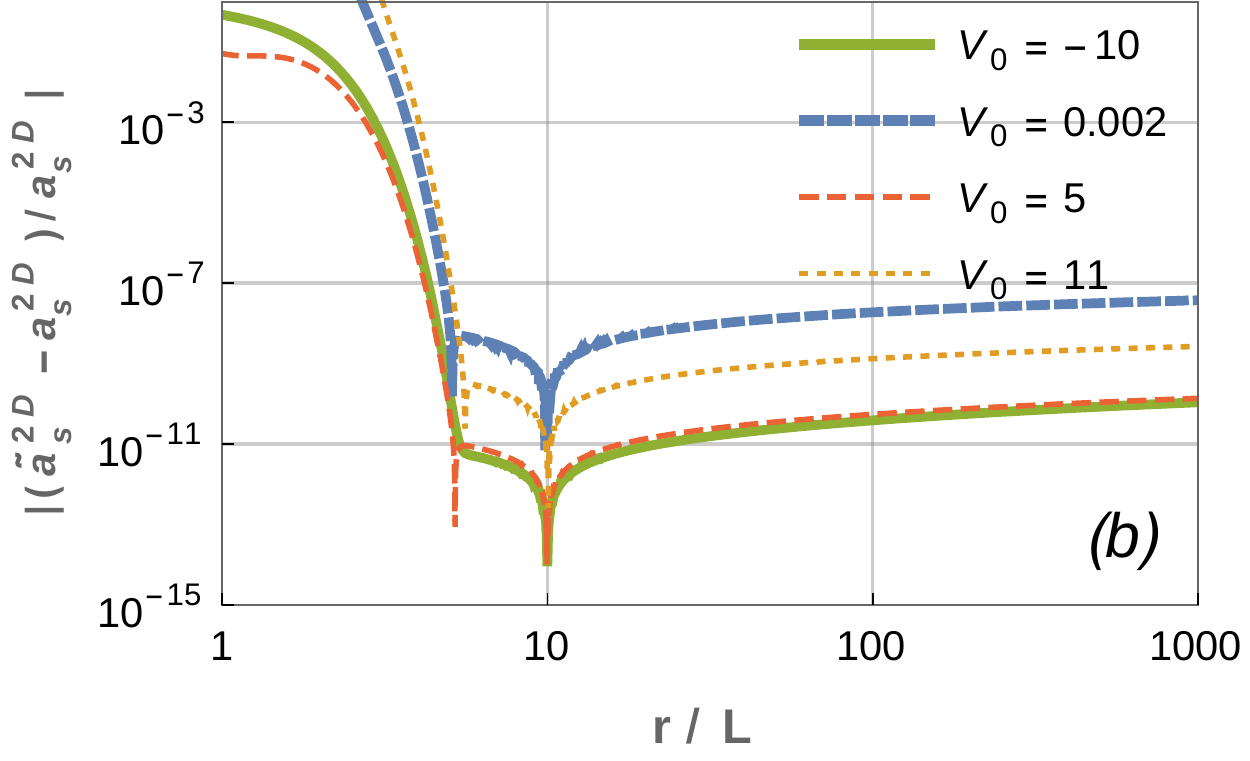}\\
\includegraphics[scale=0.5]{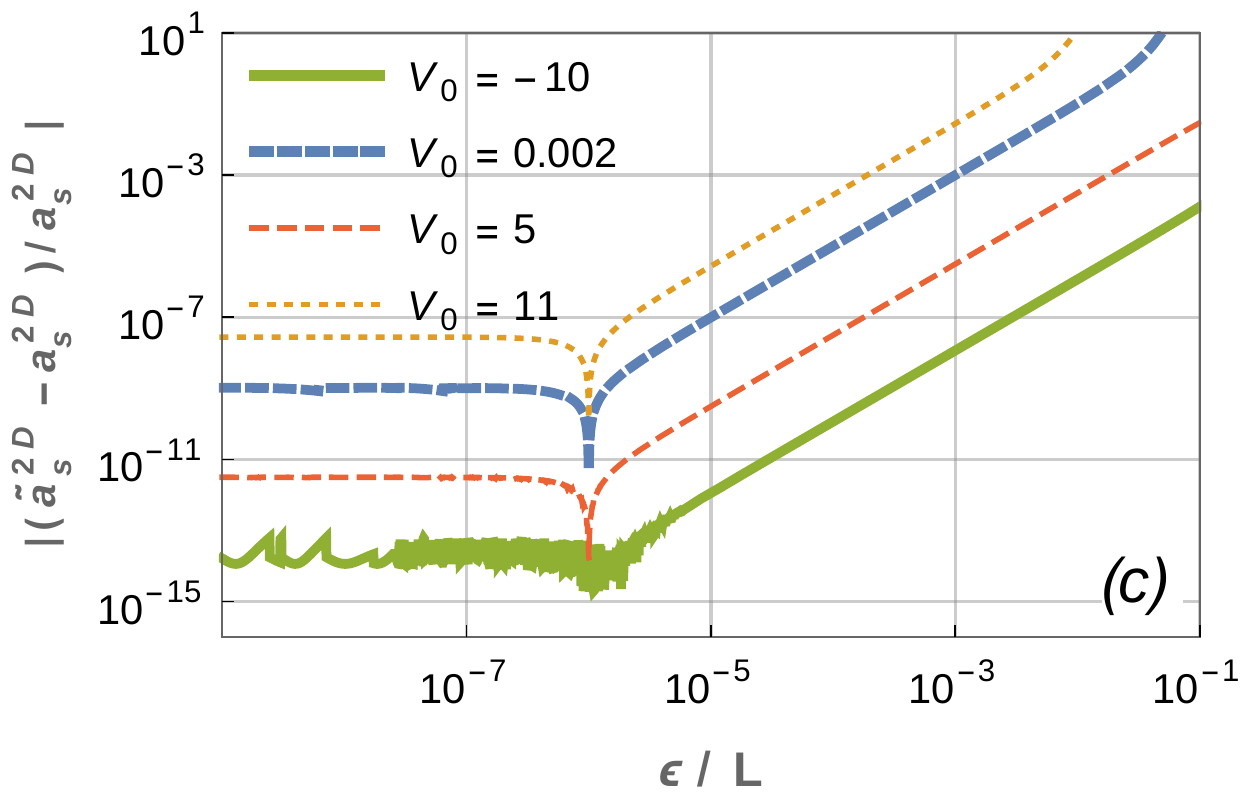}
\caption{Relative error in  the two-dimensional $s$-wave scattering length compared to a reference value $a_s^{2D}$ computed with parameters $p=11$, $r=10L$, $\epsilon=10^{-6}L$.
Shown is the parameter dependence (a) on the numerical precision $p$, (b) on the cutoff distance $r$, and (c) on the boundary parameter $\epsilon$ for different values of the potential strength $V_0$.
}
\label{fig:2Dconvg}
\end{figure}

\begin{figure}
\center
\includegraphics[scale=0.5]{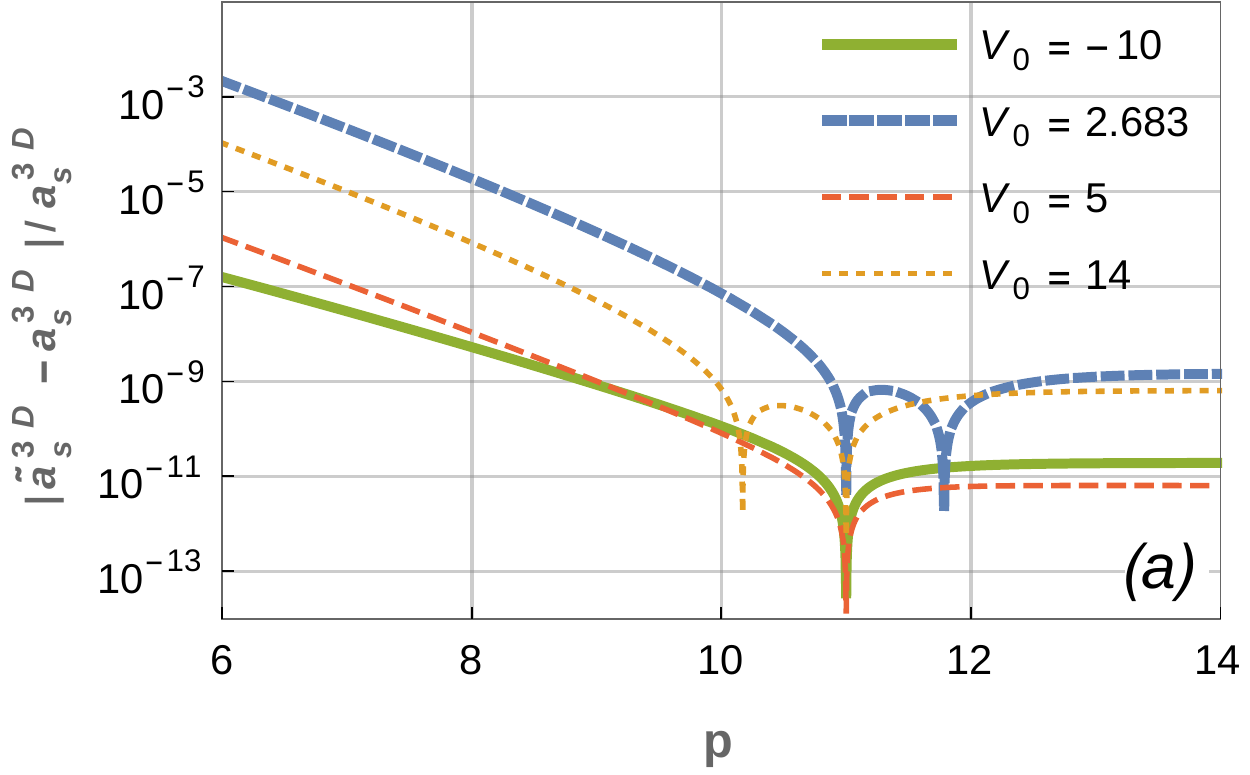} \hspace{0.5cm}
\includegraphics[scale=0.5]{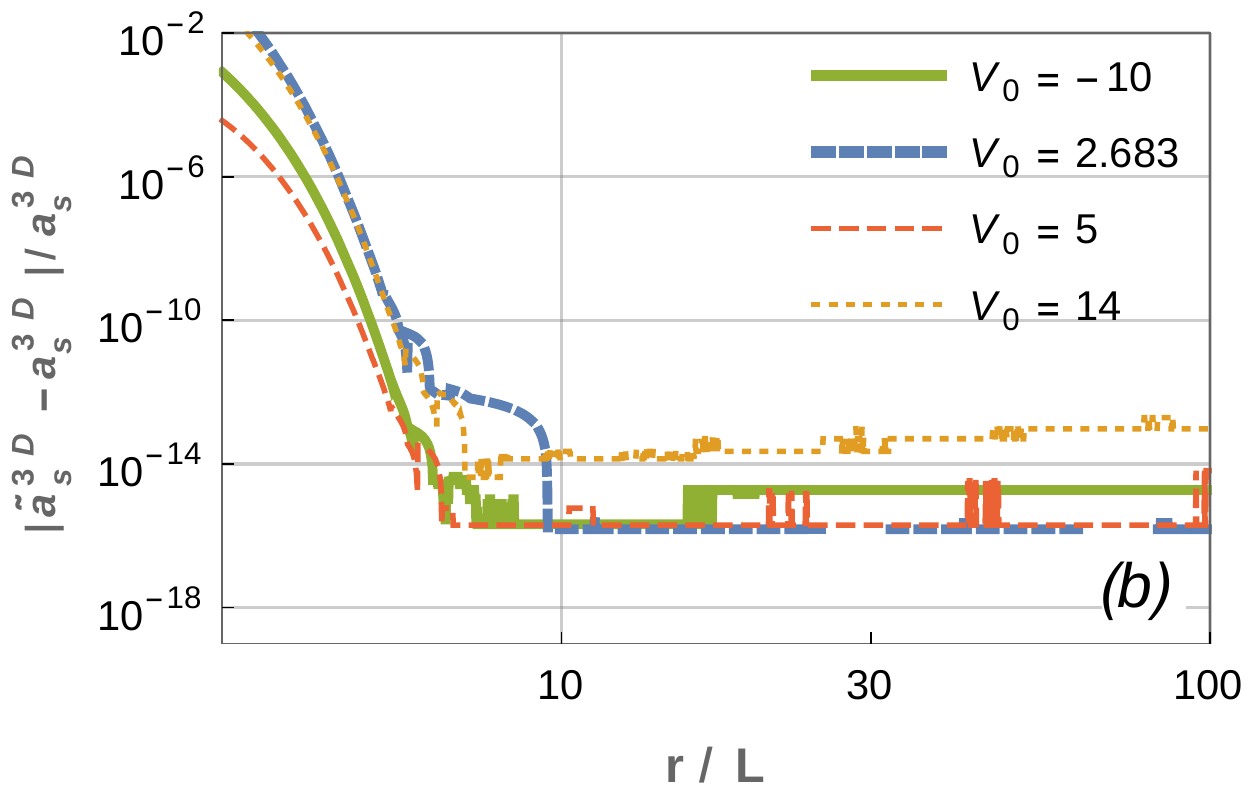}\\
\caption{Relative error in  the three-dimensional $s$-wave scattering length compared to a reference value $a_s^{3D}$ computed with parameters $p=11$, $r=10L$.
Shown is the parameter dependence (a) on the numerical precision $p$ and (b) on the cutoff distance $r$ for different values of the potential strength $V_0$.
}
\label{fig:3Dconvg}
\end{figure}

We now apply this approach to the Gaussian potential
\begin{eqnarray}
V(r)=-\frac{V_0}{2L^2} e^{-\frac{r^2}{L^2}} \ , \label{Gausspot}
\end{eqnarray}
which depends on parameters for the potential strength $V_0$ and the length scale $L$. Since we are free to use $L$ as a scale parameter, we find that the ratio $a_s/L$  depends only on the single dimensionless parameter $V_0\mu/\hbar^2$. The results of the numerical calculations and their physical interpretation will be discussed in Sec.\ \ref{sec:Approx} along with analytic approximations. Here we discuss the details and convergence properties of the numerical approach.

The numerical calculations are performed with the fourth-order Runge-Kutta
method of the Mathematica program package
 ~\cite{wolfram_research_inc._mathematica_2014}. The parameter $p$ is considered
here as a composite variable. 
We set the parameters  "AccuracyGoal" and
"PrecisionGoal", which quantify the accuracy and  precision of the numerical method, respectively, 
to the same number $p$. The "WorkingPrecision", which controls the number of the digits
in the calculations, is set to $p+5$. 
Ideally, we should consider the infinite limit
 of $p$ and $r$, and the zero limit of  $\epsilon$.
On the computer this limit is considered numerically with a finite accuracy. 
The convergence properties of the numerical procedure can be seen from Figs.\ \ref{fig:2Dconvg} and \ref{fig:3Dconvg} for the $s$-wave scattering length of the Gaussian potential in two and three dimensions, respectively. We first compute a fairly accurate reference value with a fixed choice of the accuracy parameters and then plot the relative error of the scattering length compared to the reference value as a function of the accuracy parameters.

In all cases, the  relative error  decays exponentially until 
the reference value of the accuracy parameter is reached. At  that point, due to the equality of $\tilde{a}_s^{nD}=a_s^{nD}$,
the curves
abruptly drop to zero. Beyond that point, the relative error saturates to a constant value that corresponds to the numerical error of the reference value of the
scattering length. 
In two dimensions (Fig.\ \ref{fig:2Dconvg}), the largest errors occur at $V_0=0.002 \hbar^2/\mu$ and $V_0=11 \hbar^2/ \mu$, close to divergences of the  scattering length  (see Fig.\ \ref{fig:fitting2D}).
This demonstrates how the numerical accuracy is limited near the divergences of the scattering length.  

In three dimensions, the $s$-wave scattering length diverges near to $V_0=2.683 \hbar^2/\mu$, where the  largest errors in are seen  in Fig.\ \ref{fig:3Dconvg}(a). In the same
graph, the case of $V_0=14 \hbar^2/\mu$ has the second largest numerical error. In that case the scattering length is 
close to the zero crossing ($a_s^{3D} \approx 0.05L$).
It is difficult to compute it accurately from \rref{localas3d} where a difference of small numbers [cutoff distance and inverse logarithmic derivative of $\Phi_{3D}(r)$] needs to be taken.
This effect is even more notable as a function of the cutoff distance in Fig.\ \ref{fig:3Dconvg}(b), where the error in
the case of $V_0=14 \hbar^2/\mu$ is at least one order larger at larger distances compared to other values of the
potential strength. Numerical rounding errors also explain the jumps in the cases of $V_0=-10 \hbar^2/\mu$ and $V_0=5 \hbar^2/\mu$, which are the 
limit of the chosen accuracy.


\vglue 1cm
\section{\label{sec:Approx}Approximate expressions for Gaussian potential}

\subsection{Three-dimensional case}
As can be seen in the previous section, the numerical approach is accurate in most cases, but fails near the divergences of the scattering length. Here we derive analytic approximations that can handle these numerically unstable regions. An alternative derivation based on the Lippmann-Schwinger equation is given 
in Appendix \ref{AppendixLS}.

\begin{figure}
\center
\hglue 0.5cm \includegraphics[scale=0.5]{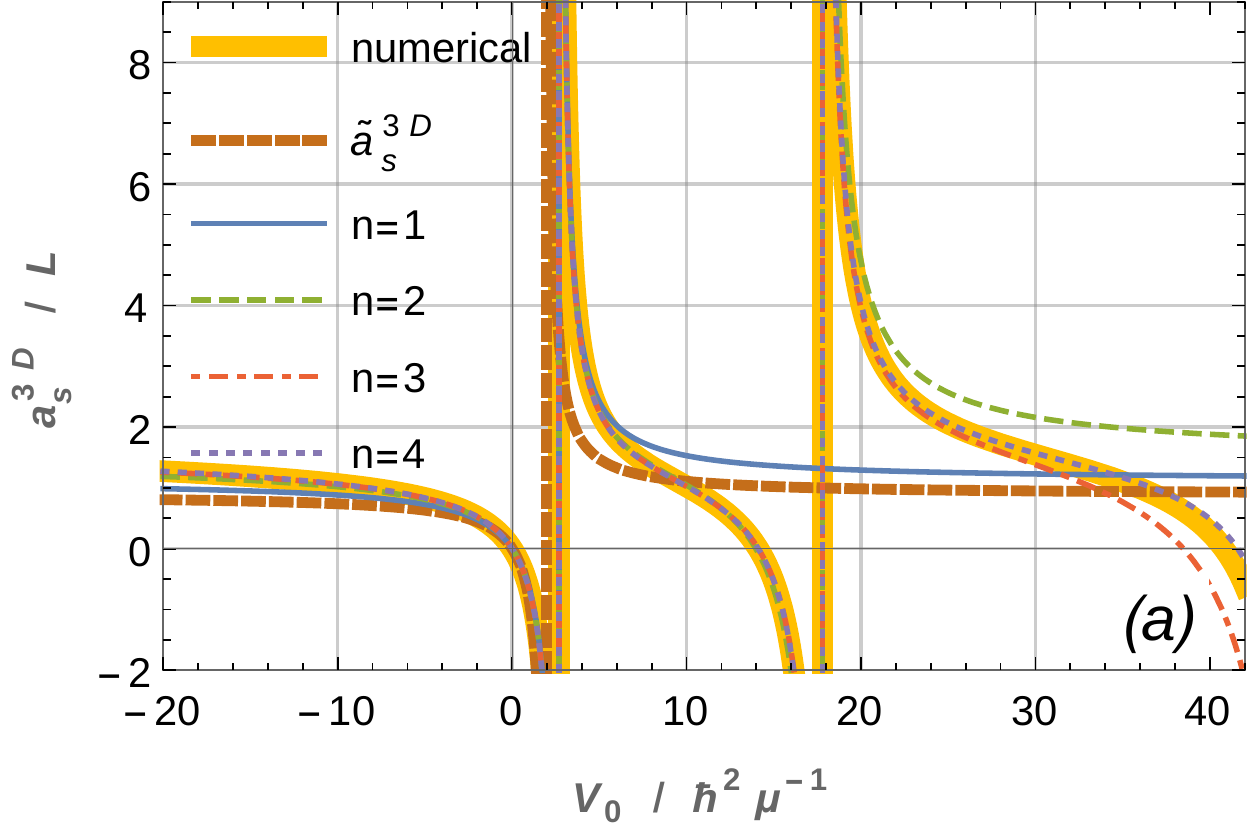} \hspace{0.5cm}
\includegraphics[scale=0.54]{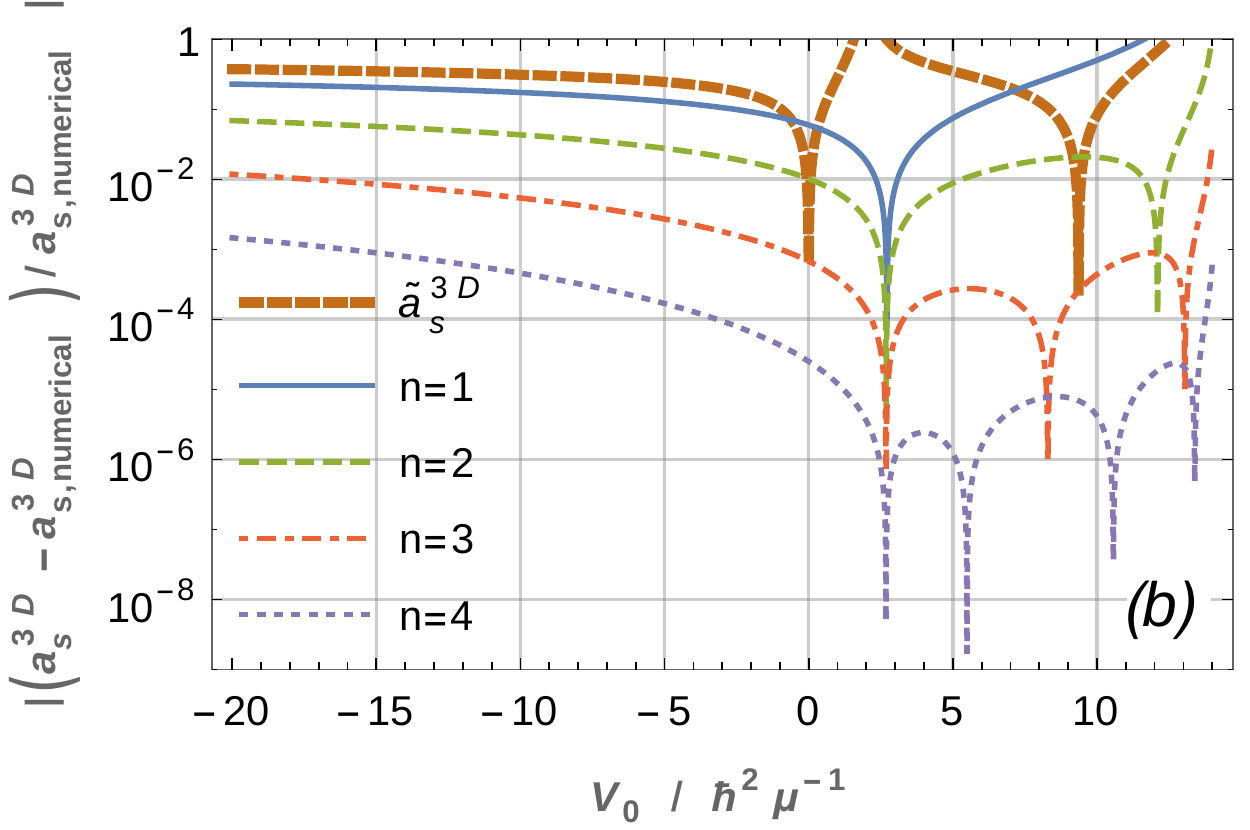}
\caption{(a) Three-dimensional $s$-wave scattering length from the numerical and the approximate expression. (b) The
 difference between the approximative and numerical scattering length values. The parameters of
the numerical simulations are set to $p=11$ and $r=10L$. The values of the parameters
for the approximative expressions can be found in Table \ref{table:fittedcoeff3d}.}
\label{fig:fitting3D}
\end{figure}

In order to  derive suitable approximations, we can make use of the fact that the Gaussian potential decays 
rapidly to zero with increasing distance. 
Contributions of the long-range part of the wave function, therefore,  become negligible when they are multiplied by this potential compared to the other parts of the Schr\"odinger equation, e.g., Eq.\ \eqref{urelsch}. Let us specifically consider the simplest case of a shallow Gaussian potential in three dimensions that has no bound states. Thus the zero-energy wave function  $u_{3D}(r)$  will be nodeless and can be safely approximated with the leading term of its Taylor expansion around $r=0$ in the product with the Gaussian potential as
\begin{eqnarray}
e^{-\frac{r^2}{L^2}} \, u_{3D}( r) & \approx& e^{-\frac{r^2}{L^2}} \, r \ .  \label{3dtaylor}
\end{eqnarray}
Note that the long-range asymptotics of  $u_{3D}(r)$ that define the scattering length are (approximately) unaffected by this procedure.
Substituting into \rref{urelsch} at $E_{rel}=0$, we  obtain the differential 
equation
\begin{eqnarray}
-\frac{\hbar^2}{2\mu} \frac{\mbox{d}^2}{\mbox{d}r^2} \bar{u}_{3D}( r) -\frac{V_0}{2L^2} e^{-\frac{r^2}{L^2}} r=0 \ . \label{approxdiffeq}
\end{eqnarray}
The function $\bar{u}_{3D}( r)$ can be obtained by integrating \rref{approxdiffeq} twice,
\begin{eqnarray}
\bar{u}_{3D}( r) &=& c_1^{3D} +c_2^{3D}r+\frac{\sqrt{\pi} V_0 \mu}{4\hbar^2} \mbox{erf} \left(\frac{r}{L}\right) \ , 
\label{sol3d}
\end{eqnarray}
where $\mathrm{erf}(x) = (2/\sqrt{\pi})\int_0^x \exp(-t^2) dt$ is the error function. The coefficients in \rref{sol3d}
can be determined by considering the boundary conditions \rrefsb{bound3d} as
\begin{eqnarray*}
c_1^{3D}=0 \ , &\hspace{2cm}&  c_2^{3D}=\frac{2-\frac{V_0\mu}{\hbar^2}}{2L} \ .
\end{eqnarray*}
Examining these wave functions in the limit when $r$ goes to infinity and using the fact that $\lim_{r \rightarrow \infty} \mbox{erf} \left(\frac{r}{L}\right) =1 $, if $L$ is finite, we obtained the following asymptotic expression:
\begin{eqnarray}
\hspace{-0.5cm} \bar{u}_{3D}( r) \approx  \frac{2-\frac{V_0 \mu}{\hbar^2}}{2}\left( \frac{r}{L} - \frac{\sqrt{\pi}}{2}\frac{V_0}{V_0-\frac{2\hbar^2}{\mu}}\right) \ , \  r \rightarrow \infty \ . \label{assymp3d}
\end{eqnarray}
Substituting \rref{assymp3d} into \rref{localas3d}, the approximate relations between 
the $s$-wave scattering length and the parameters of the potential can be found as 
\begin{eqnarray}
\frac{ \bar{a}_s^{3D}}{L} &=& \frac{\sqrt{\pi}}{2}\frac{V_0}{V_0-\frac{2\hbar^2}{\mu}}  \ . \label{as3d}
\end{eqnarray}
This approximate formula has a pole near the value of $V_0$  where the Gaussian potential well acquires the first bound state. The appearance of a pole, even though we had started out with assuming a nodeless wave function, validates the procedure but also signals a limit of validity of the approximation.
On closer inspection, we find that the expression \eqref{as3d}  describes the scattering length near the singularity qualitatively correctly,
but the position of the pole  is inaccurate. 
As can be seen in 
Fig.\ \ref{fig:fitting3D}, further singularities appear when the potential well becomes deeper and these correspond to additional bound states. Although the approximation \rrefsb{as3d} includes
 only the first singularity, it can be sufficient for the use as a pseudopotential for ultracold atoms if only the qualitative behavior of the scattering length in the presence of up to one bound state is of interest. 
 
In order to reproduce the behavior 
of the scattering length across a larger range of potential strengths, we generalize \rref{as3d} by explicitly introducing a variable number of singularities in the following way:
\begin{eqnarray}
 \frac{ a_s^{3D}}{L} &\approx& \sum_{i=1}^n \alpha_i \frac{V_0}{\left(V_0-W_i \right)} \ . \label{imprappexp3d}
\end{eqnarray}
Here, $W_i$ and $\alpha_i$ are numerically determined parameters. 
The parameters $W_i$ are set to the values of $V_0$ where the numerically determined scattering length
diverges (and changes sign). At these values of $V_0$ weakly bound states appear (see, e.g., Ref.~\cite{book:flugge99}, problem 90). To achieve a high accuracy for approximations of the $s$-wave scattering length, it is important to use accurate values for these parameters   (see the detailed description in Appendix \ref{AppendixSingularity}).
The
parameters $\alpha_i$  are obtained by nonlinear fitting
of the approximate expression \rref{imprappexp3d} to the numerically obtained scattering length. The fitting procedure is performed on the intervals $V_0 \mu/\hbar^2 \in [0,2.68] \cup [2.69,14]$ in order to avoid the singularities. 
The values of the fitted
parameters are shown in Table \ref{table:fittedcoeff3d}.
As can be seen in Fig. \ref{fig:fitting3D},
including even one additional singularity in the model 
greatly improves \rref{as3d} and qualitatively describes the
fitted region. Each
additional fitting parameter further
improves the relative accuracy by
more than one order of magnitude. In addition, the approximate
formulas also dramatically improve the
approximation of $a_s^{3D}$ outside the fitted
regime with each fitting parameter.

\begin{table}
\center
\resizebox{.5\textwidth}{!}{
\begin{tabular}{c|c|cccc}
n & $W_n \left( \hbar^2/ \mu  \right)$  & $\alpha_1$ & $\alpha_2$ & $\alpha_3$ & $\alpha_4$  \\ \hline
1 &  2.68400465092 & 1.11942413969 & & & \\
2 &  17.7956995472 & 1.12031910105 & 0.378402820446 &  & \\
3 &  45.5734799205 & 1.12034867267 & 0.322141242778 & 0.332600792963 &  \\
4 &  85.9634003809 & 1.12034897387 & 0.326461774698 & 0.135560767226 & 0.375312300726 \\ \hline
$\tilde{a}_{s}^{3D}$ &  2         & $\sqrt{\pi}/2 \approx$ 0.8862269 & & & \\

\end{tabular}}
\caption{Numerical values of parameters for the three-dimensional approximate expression \eqref{imprappexp3d}. 
}
\label{table:fittedcoeff3d}
\end{table}


\subsection{One-dimensional case}
We follow an analogous procedure to the three-dimensional case by approximately solving the Schr\"odinger equation for large $r$.
Although the form of the one-dimensional Schr\"odringer equation is equivalent to the three-dimensional one [\rref{urelsch}],
the boundary conditions of \rrefsa{bound1d} and \rrefsb{bound3d} differ. 
This has the consequence that the zeroth-order term in the Taylor expansion of $u_{1D}(r)$ does not vanish and thus we may approximate the differential equation as
\begin{eqnarray}
-\frac{\hbar^2}{2\mu} \frac{\mbox{d}^2}{\mbox{d}r^2} \bar{u}_{1D}( r) -\frac{V_0}{2L^2} e^{-\frac{r^2}{L^2}} =0 \ .
\label{1Dzeroorder}
\end{eqnarray}
Equation\ (\ref{1Dzeroorder}) can be solved and  provides
the following approximate expression for the wave function and the $s$-wave scattering length:
\begin{eqnarray}
\bar{u}_{1D}(r) &=& 1-\frac{V_0\mu}{2\hbar^2} \left[ e^{-\frac{r^2}{L^2}}+\sqrt{\pi}\mbox{erf}\left(\frac{r}{L}\right)\frac{r}{L} -1\right] \ , \label{sol1d} \\
  \frac{\bar{a}_s^{1D}}{L} &=& \ \frac{1}{\sqrt{\pi}} + \frac{2}{\sqrt{\pi}}\frac{\hbar^2}{V_0\mu}. \label{as1d} 
\end{eqnarray}
Comparing the obtained expression (\ref{as1d}) with the three-dimensional result (\ref{as3d}), it can be seen 
that the first singularity in one dimension is located in the origin, while in three dimensions it is displaced to a finite value of attractive
potential strength. As every singularity indicates the creation of a new bound state, the former statement is related
to a well-known property: in one dimension, there is always a  bound state at any nonzero attractive potential, meanwhile,
in three dimensions, the bound state appears at some finite potential strength. 

The approximate expression (\ref{as1d}) can be further improved if we expand 
the function $u_{1D}(r)$ in a Taylor series around the 
origin. As we are interested in the behavior of the singularity in the origin, we can consider the limit of $V_0$ approaching zero, 
where the coefficients of the Taylor expansion can be determined (see the detailed description in Appendix \ref{Appendix}). 
By examining the asymptotic properties of the wave function we  obtain the following approximate formula for the scattering length:
\begin{eqnarray}
\frac{\bar{\bar{a}}_s^{1D}}{L} &=&  \sqrt{\frac{2}{\pi}} + \frac{2}{\sqrt{\pi}}\frac{\hbar^2}{V_0\mu}, \label{tas1d} 
\end{eqnarray}
which differs from \rref{as1d} only in the constant offset. This expression fits better with the numerically obtained results, but it is still inaccurate at larger absolute values of the potential strength.
In analogy to the three dimensional case [\rref{imprappexp3d}], the accuracy of \rref{tas1d} can be further improved by
including additional singularities, 
\begin{eqnarray}
\frac{a_s^{1D}}{L} &\approx& \sqrt{\frac{2}{\pi}} +  \frac{2}{\sqrt{\pi}}\frac{\hbar^2}{V_0\mu} + \sum_{i=1}^n \alpha_i \frac{V_0}{(V_0-W_i)} \ ,
\label{imprappexp1d}
\end{eqnarray}
where the parameters $W_i$ are obtained directly from the numerical solution of the differential equation. 
The parameter values $\alpha_i$ are obtained nonlinearly fitting the expression \rrefsb{imprappexp1d} to the numerical data in the interval $V_0 \hbar^2/\mu \in [1.0,8.0]$. 

\begin{figure}
\center
\hglue 0.5cm \includegraphics[scale=0.5]{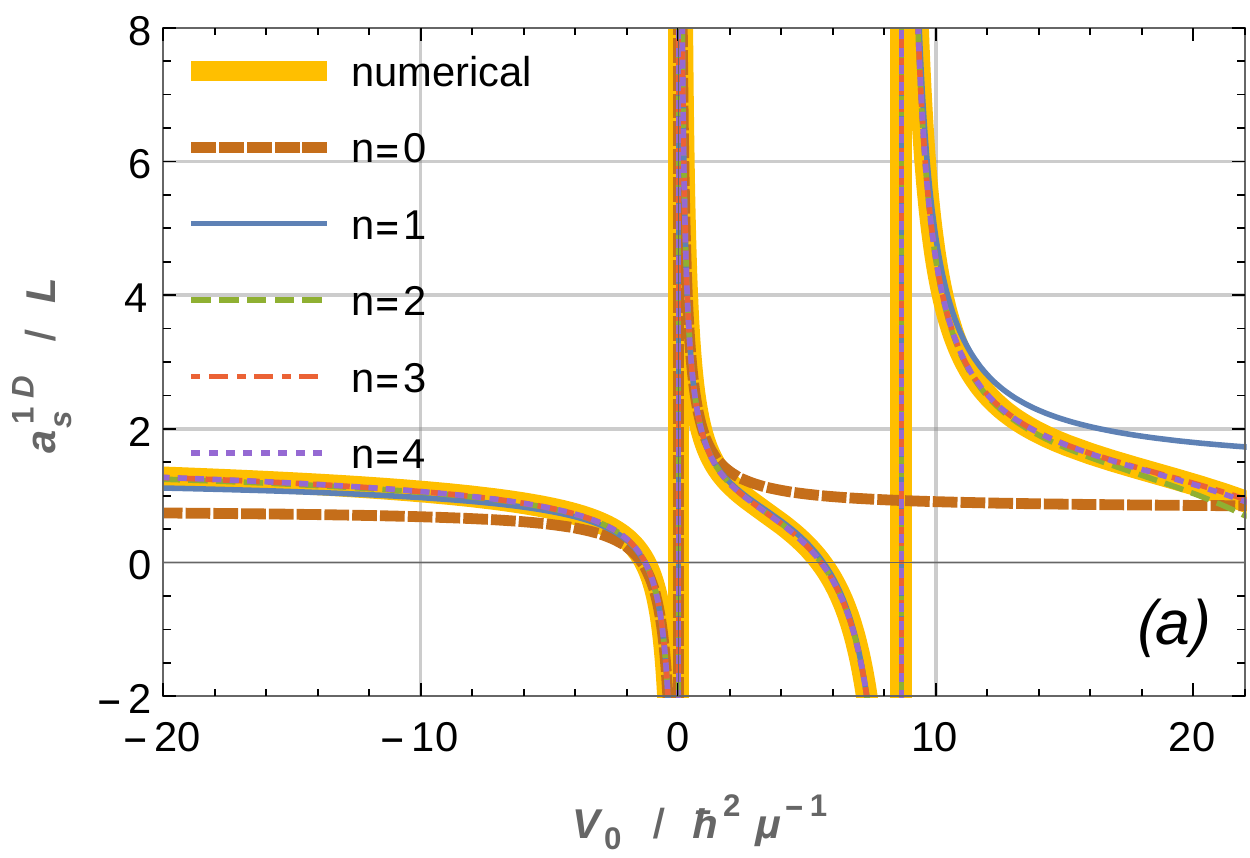} \hspace{0.5cm}
\includegraphics[scale=0.54]{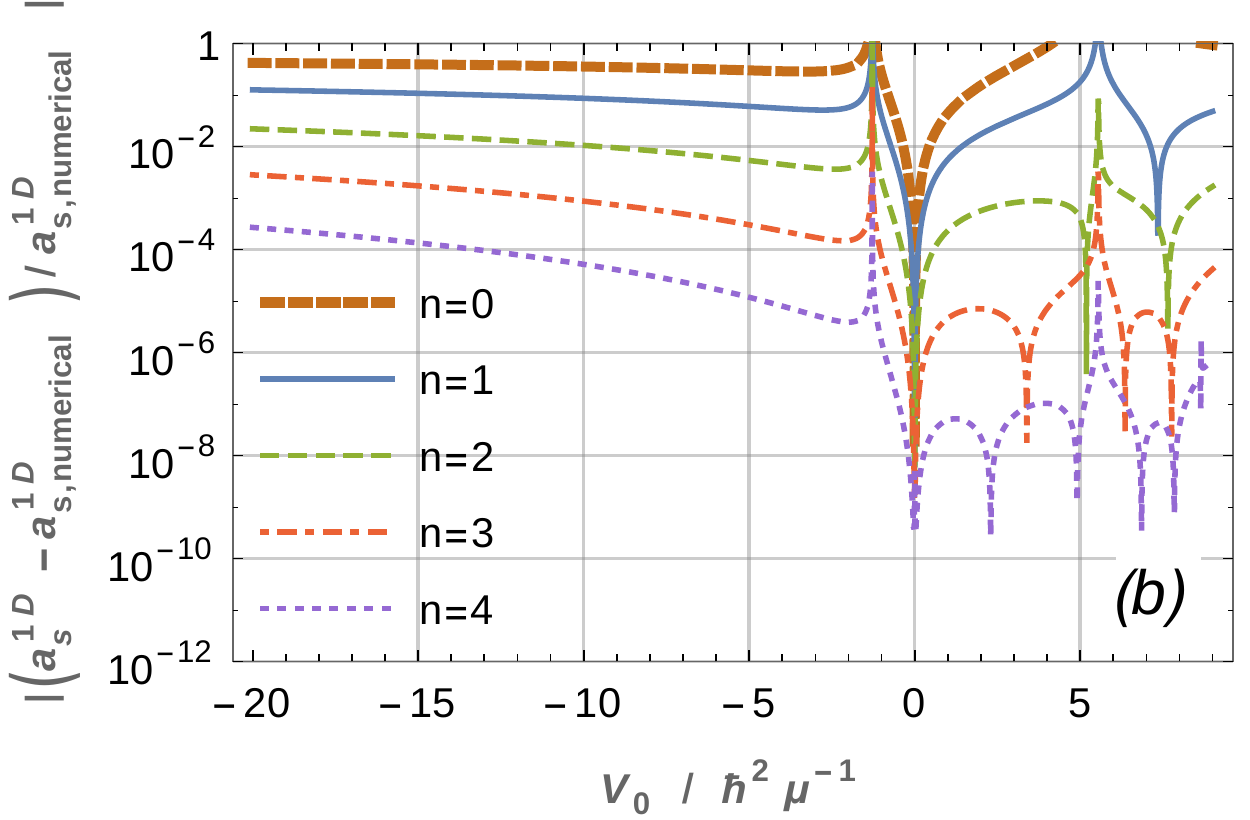}
\caption{(a) One-dimensional $s$-wave scattering length from the numerical and the approximate expression. (b) The
 difference between the approximate and numerical values of the scattering length. The parameters of
the numerical simulations are set to  $p=11$, $r=10L$. The parameter values
for the approximate expressions are tabulated in Table \ref{table:fittedcoeff1d}. The $n=0$ approximation corresponds to Eq.~\eqref{tas1d}.}
\label{fig:fitting1D}
\end{figure}

A comparison of the approximate and the numerical results is shown in Fig.\ \ref{fig:fitting1D}.
Similarly to the three-dimensional case, the relative error from the numerical solution gradually decreases with the number of
the parameter pairs. 

\begin{table}[h!]
\center
\begin{tabular}{c|c|cccc}
n & $W_n \left( \hbar^2/ \mu  \right)$  & $\alpha_1$ & $\alpha_2$ & $\alpha_3$ & $\alpha_4$  \\ \hline
1 & 8.6490975 &  0.52689372 &  &  & \\
2 & 30.106280 &  0.51419392 & 0.35899733 &  & \\
3 & 64.193333 &  0.51460375 & 0.20675606 & 0.36766012 &  \\
4 &110.88204  &  0.51459468 & 0.24033314 & 0.040512694 & 0.44420188
\end{tabular}
\caption{Numerically determined parameters for the one-dimensional approximate expression in \rref{imprappexp2d}. }
\label{table:fittedcoeff1d}
\end{table}

\subsection{Two-dimensional case}
In two dimensions the function $\Phi_{2D}(r)$ is considered, where
the corresponding Schr\"odinger equation \rrefsb{relsch} at $E_{rel}=0$ and boundary conditions \rrefsb{boundarycond}
provide the following approximate differential equation:
\begin{eqnarray}
- \frac{\hbar^2}{2 \mu } \left( \frac{\mbox{d}^2}{\mbox{d} r^2 }+ \frac{1}{r} \frac{\mbox{d}}{\mbox{d} r}  
\right) & \bar{\Phi}_{2D}(r) + V( r) = 0 \ . \label{2dapp}
\end{eqnarray}
Solving \rref{2dapp}, the radial function can be obtained as
\begin{eqnarray}
\bar{\Phi}_{2D}(r) &=& 1+\frac{V_0}{4} \left[ \mbox{Ei}\left(-r^2 \right) - \gamma -  2 \, \mbox{ln}(r) \right]\ ,
\label{sol2d}
\end{eqnarray}
where Ei$(x)=-\int_{-x}^\infty \frac{e^{-t}}{t} \mbox{d}t$ is the exponential integral function.
At large particle separation, the exponential integral function goes to zero, $\lim_{r \rightarrow \infty}  \mbox{Ei} \left(-r^2 \right)=0 $, and therefore \rref{sol2d} can be approximated with 
the following expression:
\begin{eqnarray}
\bar{\Phi}_{2D}(r) &\approx& 1-\frac{V_0\mu}{4\hbar^2} \left[ \gamma +  2 \, \mbox{ln}\left( \frac{r}{L}\right) \right] \ , \hspace{0.5cm} r \rightarrow \infty \ .
\end{eqnarray}
Using this asymptotic formula, the approximate expression of the $s$-wave scattering length can be determined from \rref{as2dnum} as
\begin{eqnarray}
\frac{ \bar{a}_s^{2D}}{L} &=& 2 e^{ \frac{-3\gamma}{2}+\frac{2\hbar^2}{V_0\mu}} \ . \label{as2d} 
\end{eqnarray}
A singularity appears again in the origin, like in the one-dimensional case, as a consequence of the fact that any arbitrarily weak Gaussian potential well in two dimensions has at least one bound state. In analogy to the procedure of Appendix \ref{Appendix}, in one dimension, we can thus determine an improved prefactor to arrive at the approximation
\begin{eqnarray}
 \frac{\bar{\bar{a}}_s^{2D}}{L} &=& \sqrt{8} e^{ \frac{-3\gamma}{2}+\frac{2\hbar^2}{V_0\mu}} \ . \label{tas2d} 
\end{eqnarray}
This approximate formula \eqref{tas2d} is 
equivalent to the previously mentioned formula of Doganov $et$ al.~\cite{doganov_two_2013}, where \rref{tas2d} was derived in a 
different  manner using perturbation theory. This expression is not very accurate at larger values of potential strength 
and can be improved by including additional singularities in the same manner as done previously to obtain
\begin{eqnarray}
\frac{a_s^{2D}}{L} &\approx& \sqrt{8} e^{-\frac{3\gamma}{2}+\frac{2\hbar^2}{V_0\mu}+
\sum\limits_{i=1}^n \alpha_i \frac{V_0}{(V_0-W_i)}} \ . \label{imprappexp2d} 
\end{eqnarray}
We determined the parameters $W_i$ with the numerical differential equation solver and fitted the parameters $\alpha_i$  on the interval $V_0 \in [1,10]\hbar^2/\mu$. 

\begin{figure}
\center
\hglue 0.5cm \includegraphics[scale=0.5]{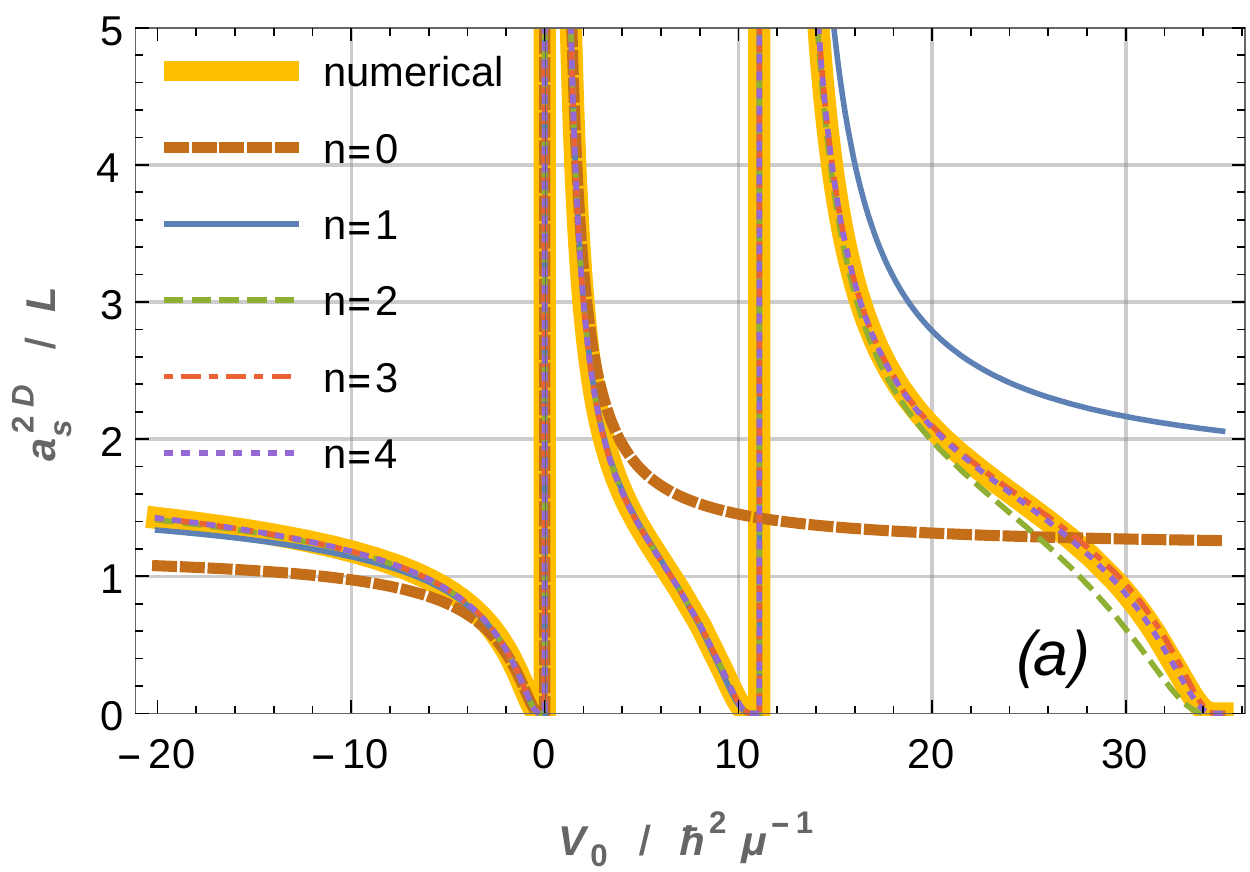} \hspace{0.5cm}
\includegraphics[scale=0.54]{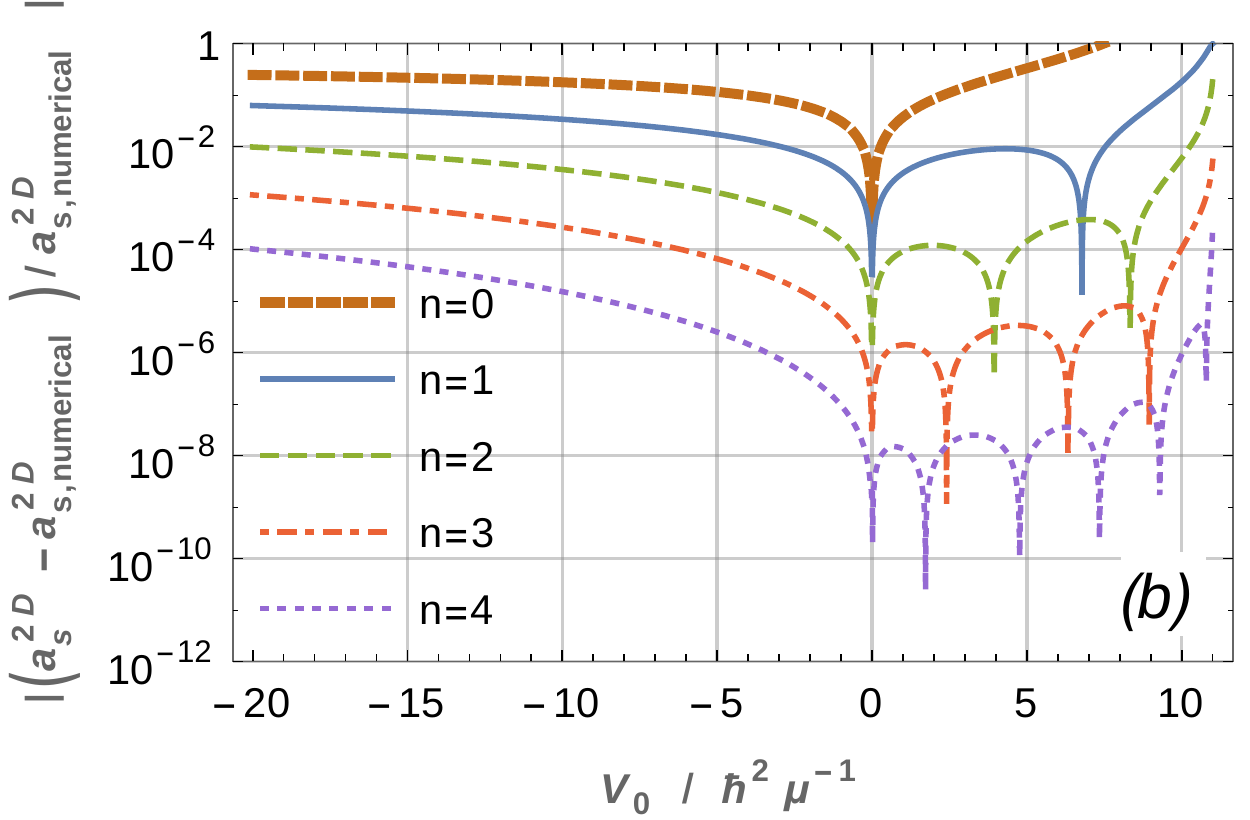}
\caption{(a) Two-dimensional $s$-wave scattering length from the numerical and the approximate expression. (b) The
 difference between the approximative and numerical scattering length values. The parameters of
the numerical simulations are set to the following: $p=11$, $r=10L$, and $\epsilon=10^{-6}L$. The values of the parameters
for the approximative expressions can be found in Table \ref{table:fittedcoeff2d}. The $n=0$ approximations correspond to Eq.~\eqref{tas2d}.}
\label{fig:fitting2D}
\end{figure}

The numerical and approximate values for the two-dimensional $s$-wave scattering length are shown in Fig.\ \ref{fig:fitting2D}. 
In contrast to the one- and three-dimensional results, the two-dimensional scattering length is always positive and single poles occur not in the scattering length itself but in its logarithm.
Similarly to the previous cases, increasing the number of fitted parameter pairs successively improves the approximate values 
for the scattering length inside and outside the fitted region.

\begin{table}
\center
\begin{tabular}{c|c|cccc}
n & $W_n \left( \hbar^2/ \mu  \right)$  & $\alpha_1$ & $\alpha_2$ & $\alpha_3$ & $\alpha_4$  \\ \hline
1 & 11.076903 & 0.33553384 &  &  & \\
2 & 35.081301 & 0.30476380 & 0.20423041 &  & \\
3 & 71.774188 & 0.30609585 & 0.10986740 & 0.19295017 &  \\
4 &121.10485  & 0.30605919 & 0.13171195 & 0.017845686 & 0.22077743
\end{tabular}
\caption{Numerically determined parameters for the two-dimensional approximate expression in \rref{imprappexp2d}. }
\label{table:fittedcoeff2d}
\end{table}

\section{Conclusion}
We have introduced approximate expressions for the $s$-wave scattering length for a Gaussian potential
in one, two, and three dimensions. These may be useful on their own or can improve the accuracy of a numerical determination of the scattering length by providing the correct asymptotic behavior near singularities. 
The lowest-level expressions can be obtained as simple parameter-free approximations derived from the two-particle 
Schr\"odinger equation. They can qualitatively describe the singularity at the first bound-state formation, 
where numerical methods usually fail or provide inaccurate answers. In one and two 
dimensions these expressions can be further improved analytically by examining  the weakly interacting limit, where the 
leading terms can be given exactly.
More accurate expressions generalize the simple formulas in a straightforward way by including additional singularities,  where the unknown parameters are determined from accurate 
numerical  computations. The obtained formulas improve the accuracy for the whole region of the potential strength. In three dimensions, where the singularity due to appearance of the first bound state appears at a finite value of the potential strength, the accuracy  of this value crucially limits the obtainable accuracy for the $s$-wave scattering length.

The Gaussian potential well has its main application for use as a pseudopotential in the description of ultracold atoms in the parameter regime between zero interaction and the first nontrivial bound state.
In this region, the relative error of the parameterized approximate formulas reaches below  $10^{-4}$ and thus they 
provide accurate, reliable, and simple formulas to connect the parameters of the Gaussian potential to the $s$-wave scattering length.

\section{Acknowledgement}
We wish to acknowledge  
Jonas Cremon, Christian Forss\'en, and Stephanie Reimann for providing a note on the numerical determination of the $s$-wave
scattering length and we thank Ali Alavi and Tal Levy for discussion and for initiating our interest in this problem. P.J. and J.B. thank the Max Planck Institute for Solid State Research for hospitality during an extended stay where this work was completed. This work was supported by the Marsden Fund of New Zealand (Contract No. MAU1604). A.Yu.Ch. acknowledges support from the JINR--IFIN-HH projects.

\appendix

\section{\label{AppendixBC}Boundary conditions for the scattering problem}



We  consider a short-range interaction potential as specified in Sec.~\ref{sec:2bp} and extend a similar argument given in Ref.\ \cite{book:perelomov60} (see also \cite{cherny_dilute_2001} for the boundary conditions in two dimensions) to subleading order.
To obtain the boundary conditions, let us multiply \rref{relsch} by $r^2$ and 
consider the limit of $r \rightarrow 0$ as
\begin{eqnarray}
\lim\limits_{r \rightarrow 0} {\Bigg[ }
-\frac{\hbar^2}{2\mu}\left( r^2 \frac{\mbox{d}^2}{\mbox{d}r^2} + r(n-1)
\frac{\mbox{d}}{\mbox{d}r}  \right)\Phi_{nD}( r) \, + \hspace{0.5cm}& 
\label{rsqrelsch} \\
+ r^2\left[ V(r)- E \right] \Phi_{nD}( r) {\Bigg ]}=0 \ .& \nonumber
\end{eqnarray}
The second term vanishes more quickly than the other ones
in the limit of $r \rightarrow 0$ if 
the potential is nonsingular at the origin and diverges, at most, like $1/r^s$ with $s<2$.
In order to obtain the leading term of the radial wave function, we may
ignore this term and
consider the  differential equation
\begin{eqnarray}
\hspace{-1cm} \lim\limits_{r \rightarrow 0} {\Bigg[ } -\frac{\hbar^2}{2\mu}\left( r^2 \frac{\mbox{d}^2}{\mbox{d}r^2} + r(n-1)
\frac{\mbox{d}}{\mbox{d}r}  \right)\Phi^a_{nD}( r)  {\Bigg ]}=0 \ , \label{assympeq}
\end{eqnarray}
where $\Phi^a_{nD}( r)$ is asymptotic solution of $\Phi_{nD}( r)$ in the origin. 

Let us first  consider the cases of two and three dimensions. Here the origin is a singular point due to the $1/r$ term in Eq.\ \eqref{relsch}. For  Eq.\ \eqref{assympeq}, we find the explicit solutions
\begin{eqnarray}
\Phi^a_{2D}( r) &=& d_{2D}\, \mbox{ln}(r) + c_{2D} \ , \label{assymptotic2d}\\
\Phi^a_{3D}( r) &=& \frac{d_{3D}}{r}+c_{3D} \ . \label{assymptotic3d}
\end{eqnarray}
The parameters $d_{2D}$, $c_{2D}$, $d_{3D}$, and $c_{3D}$ are arbitrary constants. The functions $\mbox{ln}(r)$ and $1/r$ are  
singular at $r=0$, which would provide a nonsmooth wave function. 
Moreover, in the kinetic part of the Hamiltonian, the Laplacian operator would generate a Dirac $\delta$ contribution that is not compensated by the potential part 
in the Schr\"odinger equation \cite{book:perelomov60}. 
Hence, these irregular parts should be eliminated by setting the 
scalar factors $d_{2D}$ and $d_{3D}$ to 0. Setting the remaining 
coefficients $c_{nD,1}$ and $c_{nD,2}$ to 1, 
we obtain the boundary condition $\Phi(0) = 0$, i.e., the first part of Eq.\ \eqref{eq:bcs}.

In one dimension the origin is a regular point of Eq.\ \eqref{relsch} and parity is a good quantum number. An explicit solution is
\begin{align}
\Phi^a_{1D}( r) = d_{1D}{r} + c_{1D} \ . \label{assymptotic3d}
\end{align}
Since $s$-wave scattering demands an even-parity solution, we set $d_{1D}=0$ and $c_{1D}=1$ to obtain the boundary conditions \eqref{eq:bcs} in the one-dimensional case.

It remains to derive the correct boundary condition for the first derivative, $\Phi'_{nD}(0)$. A  separate consideration is necessary here, since the sub-leading term in the radial wave function may lead to a divergent derivative.
We specifically consider the case $E=0$ but nonzero energies can be studied the same way. Let us integrate the Schr\"odinger equation in $n$ dimensions,
\begin{align}\label{eq:sch_nD}
\nabla^2\Phi(x)= \frac{2\mu}{\hbar^2}V(x)\Phi(x)
\end{align}
over a small ball of radius $r$. For the left-hand side, one can apply  Gauss' theorem, which yields for a radially symmetric wave function,
\begin{align}\label{eq:Gauss}
\int_{|\vec{x}|\leqslant r}\nabla^2\Phi(x)\mbox{d}^n x= \oint_{|\vec{x}|=r}\mbox{d}\vec{S}\cdot\nabla\Phi(x)=S_n\Phi'(r)r^{n-1},
\end{align}
where $S_n=2\pi^{n/2}/\Gamma(n/2)$ is the surface area of the sphere of the 
unit radius in $n$ dimensions, and 
$\Gamma(z)$ denotes the Gamma function. When $\Phi(0)=1$, the leading term on 
the right-hand side of \eqref{eq:sch_nD} becomes $\frac{2\mu C}{\hbar^2}r^{-s}$, where $C$ is a constant. After integration, this term yields
\begin{align}
\int_{|\vec{x}|\leqslant r}\frac{2\mu}{\hbar^2}V(x)\Phi(x)\mbox{d}^n x& \simeq \int_0^r \frac{2\mu C}{\hbar^2}x^{-s}S_n x^{n-1}\mbox{d} x \nonumber\\
&\simeq S_n\frac{2\mu C}{\hbar^2}\frac{r^{n-s}}{n-s}.\label{eq:Vrphi}
\end{align}
Comparing the above equations yields
\begin{align}\label{eq:phipr}
\Phi'(r)\simeq \frac{2\mu C}{\hbar^2}\frac{r^{1-s}}{n-s},
\end{align}
which leads to the boundary condition $\Phi'(0)=0$ if $s<1$ in all three dimensions.

In summary, we have derived the boundary conditions \eqref{eq:bcs} for potentials that are regular or divergent with a leading divergence at the origin $\propto 1/r^s$ with $s<1$. This justifies the numerical procedure discussed in Sec.\ \ref{sec:Numerical}. If $1<s<2$, the radial wave function still takes a finite value at the origin, but the first derivative diverges. In this case, the numerical procedure will have to be modified.

%
%
%


\section{\label{AppendixSingularity}Accuracy of the $s$-wave scattering length in three dimensions at the singularity}
\begin{figure}
\center
\hglue 0.25cm \includegraphics[scale=0.44]{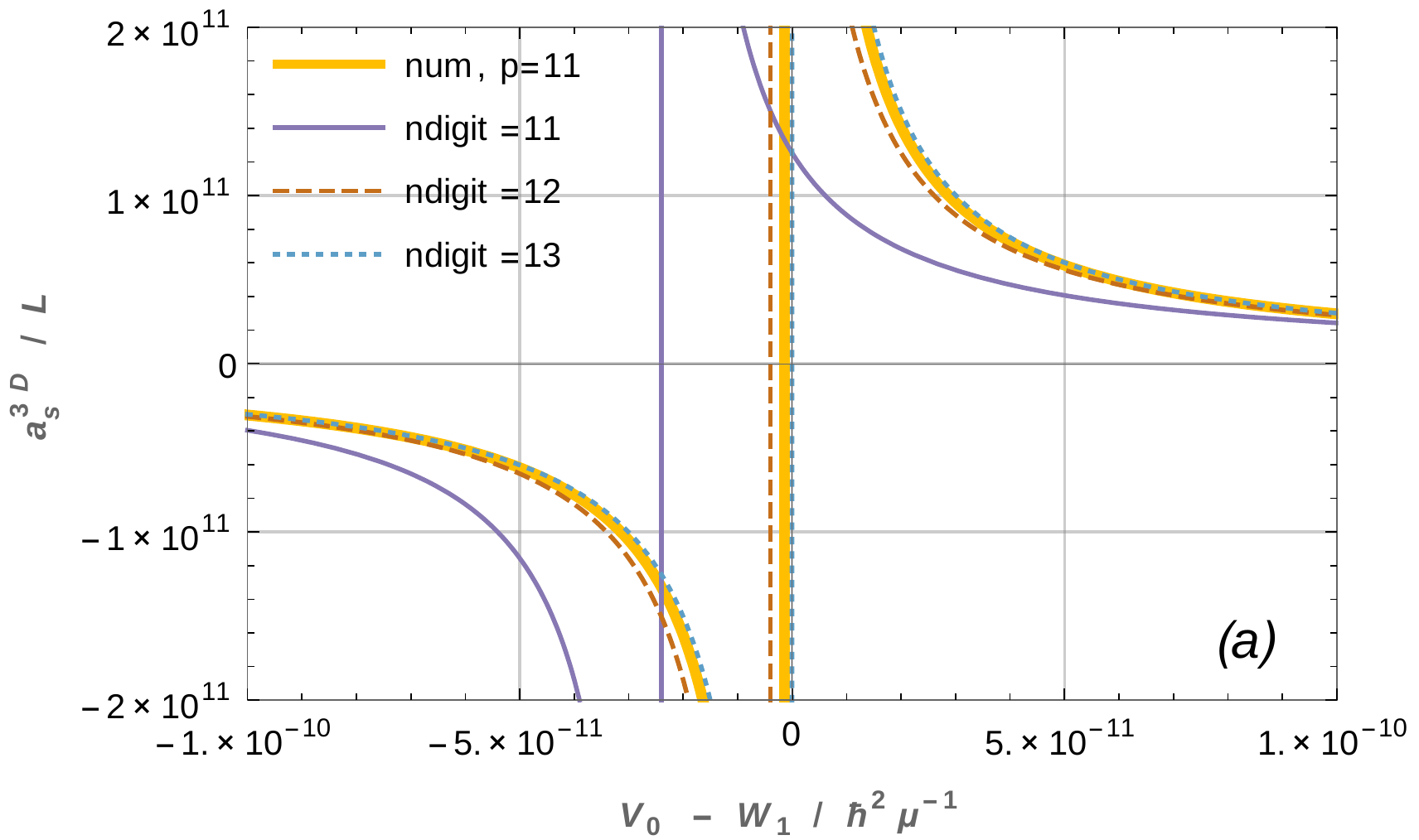} \hspace{0.5cm}
\includegraphics[scale=0.48]{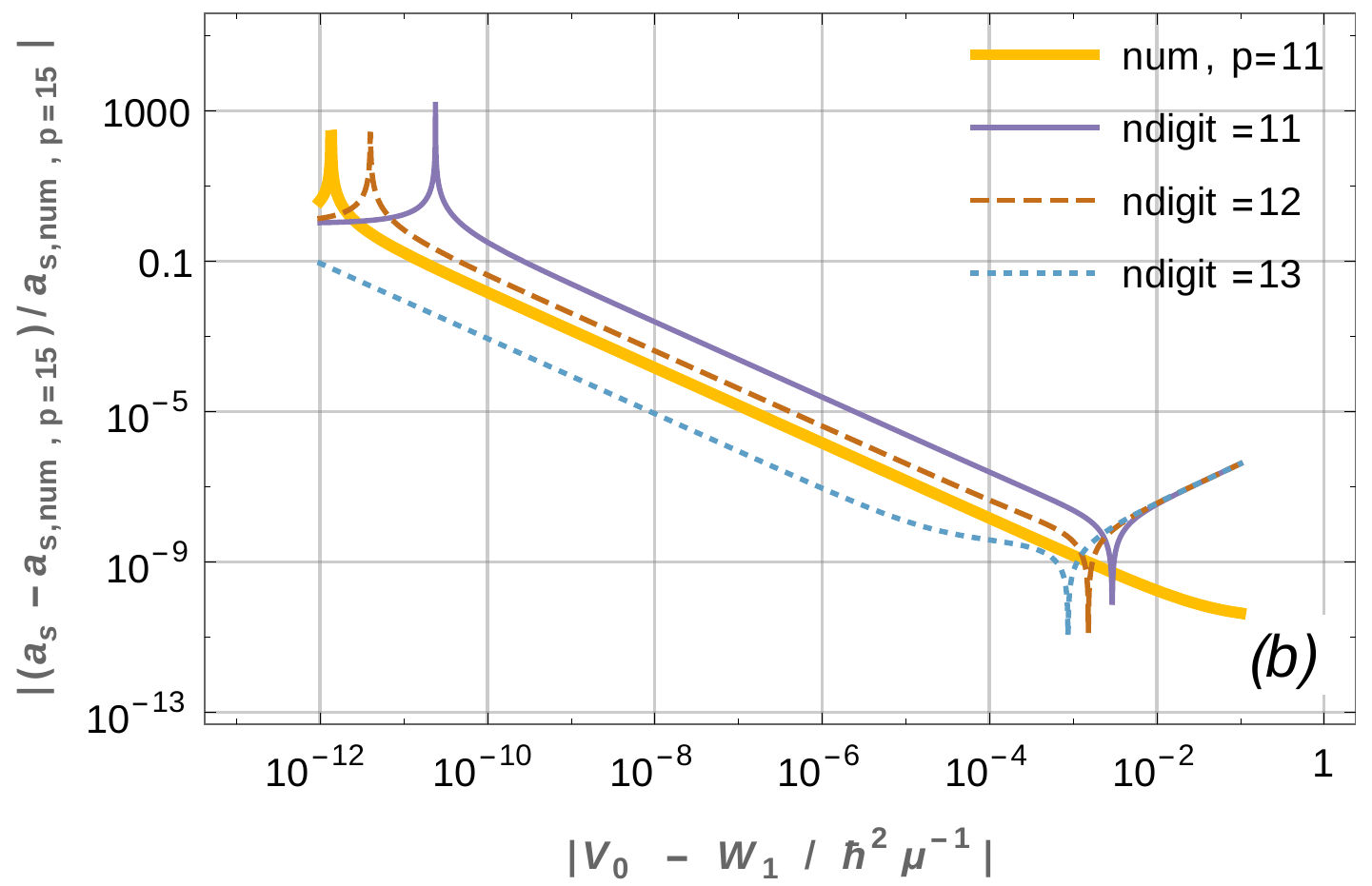}
\includegraphics[scale=0.5]{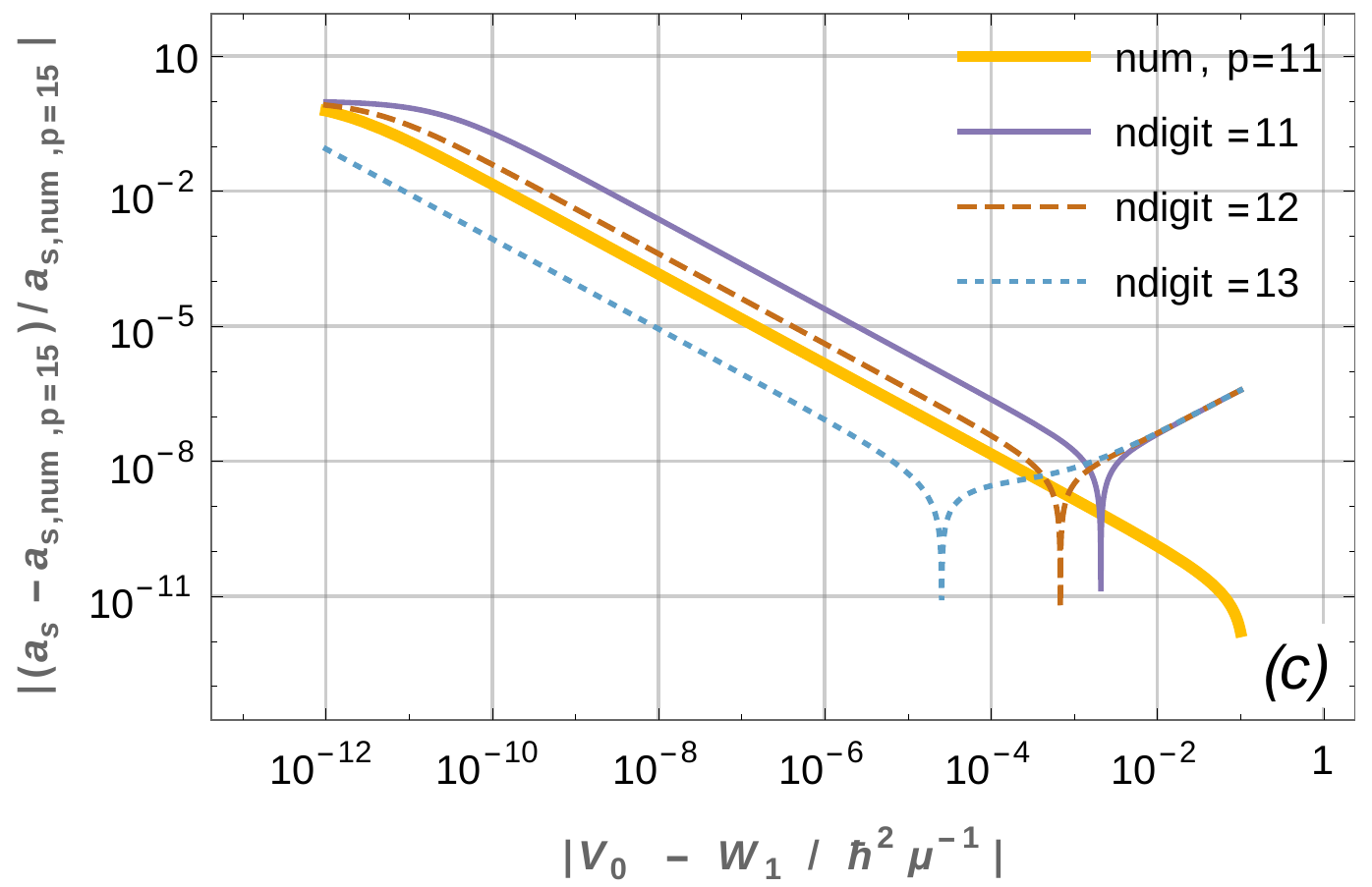}
\caption{Importance of the location of the first pole on the accuracy of the scattering length in three dimensions. (a) $s$-wave scattering length from the numerical and the approximate expressions. 
In the approximate expressions, the scattering length is calculated with varying precision ($\mathrm{ndigit}$ decimal digits) of the position of the first singularity $W_1$. (b), (c) The
errors of the approximate and numerical values of the scattering length. (b) 
$V_0 < W_1$, (c) $V_0 > W_1$ . 
Reference values for determining the zero points of the $x$ and $y$ axes, respectively, are calculated by numerical calculations with a high level of accuracy ($p=15$ for $a_s$; reference value $W_1= 2.684004650924 \hbar^2/\mu$). }
\label{fig:3Dsing}
\end{figure}

In three dimensions the accuracy of the $s$-wave scattering length is limited  mainly by the accuracy of the position $W_1$ of the first singularity. We determine this position 
using the numerical differential equation solver by finding the value of $V_0$ where the scattering length changes sign. The accuracy of
this position can be checked by increasing the accuracy of the calculation itself. We found that  $W_1$ can be determined with very good accuracy of 12 digits when $p=11$. 

The value of $W_1$ can, in principle, also be obtained by diagonalizing the Hamiltonian in a plane-wave basis and determining the value of $V_0$ where the 
ground-state energy crosses zero, extrapolating to the limits of infinite box size and basis set. We found a value that is consistent with the result from the differential equation solver to three digits of accuracy, but were not able to reach higher accuracy with the diagonalization procedure due to limitations of the extrapolation procedures. Thus we have used values extracted from the differential equation solver for the numerical results presented in this paper.

The $s$-wave scattering length is plotted with different accuracy of $W_1$  in Fig.\ \ref{fig:3Dsing}. The parameters $\alpha_i$ and the 
$W_j$ ($j>1$) are set according to Table \ref{table:fittedcoeff3d} and are kept unchanged. 
The poles with the minimums on the error curves correspond to the crossing of the reference curve. The poles with the maximums come from 
the inaccurate position of the singularity. Increasing the accuracy of $W_1$ significantly improves $a_s^{3D}$ as well. In the
main part of the paper $\mathrm{ndigit}=12$ decimal digits of accuracy are used for $W_1$, where the relative error is below $10^{-5}$, if the potential
strength is within $W_1 - 10^{-6} \hbar^2/\mu< V_0 < W_1 + 10^{-6} \hbar^2/\mu$.


\section{\label{AppendixLS}Alternative derivation of the approximate expressions for the $s$-wave scattering length}
\subsection{Three-dimensional case}
Let us introduce dimensionless variables
\begin{align}
  y&=r/L,  \label{yvar}\\
  \eta &=V_0\mu/\hbar^2, \label{etavar}
\end{align}
with which the Schr\"odinger equation (\ref{relsch}) can be written in the following form:
\begin{align}\label{schr_dim}
\left( \frac{\mbox{d}^2}{\mbox{d}y^2} + 2{y}
\frac{\mbox{d}}{\mbox{d}y} + \eta \exp(-y^2) \right)\tilde{\Phi}_{3D}(y)=E \tilde{\Phi}_{3D}(y) \ .
\end{align}
One can see from the definitions of the scattering length (\ref{assPhy}) and Eqs.~(\ref{yvar}) and (\ref{schr_dim}) that
the ratio $a_s/L$  depends only on the single dimensionless parameter $\eta$. In the following, let us consider the $E=0$ case in order to
determine the $s$-wave scattering length.

The Schr\"odinger equation (\ref{schr_dim}) can be transformed [with $\tilde{u}_{3D}(y) = y \tilde{\Phi}_{3D}(y)$] to the Lippmann-Schwinger equation~\cite{lippmann50},
\begin{align}
\tilde{u}_{3D}(y)=y-\eta\int\limits_0^{y} \mbox{d} x\, (y-x)\exp(-x^2)\tilde{u}_{3D}(x) \ .
\label{LS3D}
\end{align}

The $s$-wave scattering length can be expressed with a simple form if we substitute \rrefsa{yvar} and \rrefsb{LS3D} into \rref{assPhy},
\begin{align}
 {a}_s^{3D}/{L}={c_2}/{(c_1+1)} \ , \label{asym3D1} 
\end{align}
where
\begin{align}
 c_1&=-\eta\int\limits_0^{\infty}\mbox{d} x\, \exp(-x^2)\tilde{u}_{3D}(x) \ , \label{c1} \\
 c_2&= -\eta \int\limits_0^{\infty}\mbox{d} x\,x\exp(-x^2)\tilde{u}_{3D}(x) \ .\label{c2}
\end{align}

Let us solve Eq.~(\ref{LS3D}) with iterations. In the first step, we consider $\eta=0$ on the right hand side,
\begin{eqnarray*}
\tilde{u}^{(0)}_{3D}(y)=y \ .
\end{eqnarray*} 
Substituting it into \rrefsa{asym3D1}-\rrefsb{c2} the zero-order approximation for the scattering length can be obtained,
\begin{align}
\frac{ \bar{a}_s^{3D}}{L} =-\frac{\sqrt{\pi}}{4}\frac{\eta}{1-\frac{\eta}{2}}  \ , \label{as3dLS}
\end{align}
which is equivalent to the analytical expression \rrefsb{as3d} in the main text.


\subsection{One-dimensional case}

In one dimension, the Lippmann-Schwinger equation and the approximate expression of the scattering length are derived analogously to the three-dimensional case, 
\begin{align}
\tilde{u}_{1D}(y)&=1-\eta\int\limits_0^{y}\mbox{d} x\, (y-x)\exp(-x^2)\tilde{u}_{1D}(x) \ , 
\label{LS1D} \\
 {a}_s^{1D}/{L}&={(c_2-1)}/{c_1} \ , \label{asym1D1}
\end{align}
with the same relations (\ref{c1}) and (\ref{c2}) for the constants $c_1$ and $c_2$ as in the three-dimensional case.
The  difference from the three-dimensional solution arises from the different boundary conditions \rrefsb{bound1d} and \rrefsb{bound3d}.

We solve \rref{LS1D} iteratively. In the first step we consider $\tilde{u}^{(0)}_{1D}(y)=1$, from which the first-order wave function and zero-order scattering length can be 
obtained,
\begin{align}
\bar{u}_{1D}(y)&=1-\frac{\eta}{2} \left[ e^{-y^2}+\sqrt{\pi}y\erf(y)-1\right],\label{sol1dLS} \\
\frac{ \bar{a}_s^{1D}}{L} &=\frac{2}{\sqrt{\pi}}\frac{1}{\eta}+\frac{1}{\sqrt{\pi}}. \label{as1dLS}
\end{align}
The first-order scattering length, obtained with Eq.~(\ref{sol1dLS}), is given by
\begin{eqnarray}
\frac{\bar{\bar{a}}_s^{1D}}{L} &=&  \frac{2}{\sqrt{\pi}}\frac{1}{\eta}+\sqrt{\frac{2}{\pi}} + O(\eta) \ . \label{tas1d_app}
\end{eqnarray}
The zeroth- and first-order term recover Eqs.\ \eqref{as1d} and \eqref{tas1d} from the main text.

\subsection{Two-dimensional case}

In two dimensions, the Lippmann-Schwinger equation for the function $\tilde{\Phi}_{2D}(y)$, obeying the Schr\"odinger equation (\ref{schr_dim}), takes the form \cite{khuri09}
\begin{align}
\tilde{\Phi}_{2D}(y)=1-\eta\int\limits_0^{y}\mbox{d} x\,x \ln(y/x)\exp(-x^2)\tilde{\Phi}_{2D}(x) \ .
\label{LS2D}
\end{align}
Comparing its long-range asymptotics with the definition (\ref{assPhy}) and using Eq.~(\ref{yvar}), we derive
\begin{equation}
 {{a}_s^{2D}}/{L}=e^{\frac{c_1-1}{c_2}-\gamma+\ln 2},\label{asym2D1}
\end{equation}
where
\begin{align}
 c_1&=-\eta\int_0^{+\infty}\mbox{d} x\, \ln(x)\exp(-x^2)\tilde{\Phi}_{2D}(x) \ , \label{c12d} \\
 c_2&= -\eta \int_0^{+\infty}\mbox{d} x\,x\exp(-x^2)\tilde{\Phi}_{2D}(x) \ . \label{c22d}
\end{align}
Similar to the previous sections, the zero-order approximation for the scattering length is obtained with the zero-order function $\tilde{\Phi}^{(0)}_{2D}(x)=1$,
\begin{eqnarray}
\bar{a}_s^{2D}/{L} &=& 2 e^{ \frac{-3\gamma}{2}+\frac{2}{\eta}} \ . \label{as2dLS}
\end{eqnarray}
The first-order wave function, obtained with the first iteration, is given by
\begin{eqnarray}
\bar{\tilde{\Phi}}_{2D}(y) &=& 1-\frac{\eta}{4} \left[ \gamma + 2 \ln(y) -\mbox{Ei}\left(-y^2 \right)\right]\ ,
\label{sol2d}
\end{eqnarray}
where Ei$(z)=-\int_{-z}^\infty \frac{e^{-t}}{t} \mbox{d}t$ is the exponential integral function. Substituting it into Eqs.~(\ref{c12d}) and (\ref{c22d}), and using (\ref{asym2D1}), gives us
\begin{eqnarray}
 {\bar{\bar{a}}_s^{2D}}/{L} &=& \sqrt{8} e^{ \frac{-3\gamma}{2}+\frac{2}{\eta}+O(\eta)} \ . \label{tas2dLS}
\end{eqnarray}
As can be seen the obtained \rrefsa{as2dLS} and \rrefsb{tas2dLS} are equivalent to \rrefsa{as2d} and \rrefsb{tas2d} from the main text.

\section{\label{Appendix}Derivation of approximate formula (\ref{tas1d}) for the one-dimensional $s$-wave scattering length}

As we previously discussed in Appendix \ref{AppendixBC}, the one-dimensional wave function $u_{1D}(r)$ is even,
hence, its power-series expansion can be written in the following form:
\begin{eqnarray}
u_{1D}(r)=\sum_{k=1}^{\infty} b_k r^{2k} \ . \label{expansionu}
\end{eqnarray}
Substituting back \rref{expansionu} into \rref{urelsch}, we got the following differential equation:
\begin{eqnarray}
u_{1D}''(r)=-V_0e^{-r^2} \sum_{k=1}^{\infty} b_k r^{2k} \ , \label{1dschinfo}
\end{eqnarray}
where $ b_0$ is chosen to be one due to the boundary condition \rrefsb{bound1d}. Using the usual Taylor-expansion identity as
\begin{eqnarray*}
 b_k= \frac{1}{(2k)!}\left. \frac{\mbox{d}^{2k} u_{1D}(r)}{\mbox{d}r^{2k}} \right|_{r=0} \ ,
\end{eqnarray*}
the parameters can be determined from \rref{1dschinfo} or its corresponding derivative form as
\begin{align*}
 b_k=-\frac{V_0}{(2k)!} \sum_{l=0}^{k-1} {\Bigg [} & \binom{2k-2}{2l}   \cdot \\ 
& \cdot \left( 2k -2 -2l\right)!   \ b_{k-l-1} \left. \frac{\mbox{d}^{2k}}{\mbox{d}r^{2k}}e^{-r^2} \right|_{r=0} 
{\Bigg ]}\ .
\end{align*}
Using the following identities of the Hermite polynomials:
\begin{eqnarray*}
\left. \frac{\mbox{d}^{2k}}{\mbox{d}r^{2k}}e^{-r^2} \right|_{r=0}={\mathcal H}_{2k}(0)=(-1)^k\frac{(2k)!}{k!} \ ,
\end{eqnarray*}
parameter $ b_k$ can be expressed as a linear combination of $ b_m$ ($m < k$) as
\begin{eqnarray}
 b_k=\frac{V_0}{2k(2k-1)} \sum_{l=0}^{k-1} (-1)^{l+1} \frac{ b_{k-1-l}}{l!} \ . \label{alphaeq}
\end{eqnarray}
With \rref{alphaeq}, all the $ b_k$ can be determined, hence function $u_{1D}(r)$ can be given explicitly in the 
power-series form. 

In order to determine the $s$-wave scattering length, function $u_{1D}(r)$ should be examined in the asymptotic limit $r \rightarrow \infty$, which 
is difficult to handle in \rref{expansionu}. However, a different form of 
$u_{1D}(r)$ can be considered as
\begin{align}
u_{1D}(r)=&1+V_0 c_0- \label{otheransatz} \\ 
&-V_0 \left( e^{-r^2}\sum_{k=0}^{\infty} c_k r^{2k}+d \sqrt{\pi} \, r \, \mbox{erf}\left(r\right) \right) 
\ , \nonumber
\end{align}
where $c_k$ and $d$ are real coefficients. Equation (\ref{otheransatz}) satisfies the Schr\"odinger equation 
(\ref{urelsch}), if the coefficients  $c_k$ and $d$ are chosen properly. We can make a relation 
between \rrefsa{1dschinfo} and \rrefsb{otheransatz} by expanding in Taylor series of \rref{otheransatz}, where
we obtain the following relations:
\begin{eqnarray}
 b_0 &=& -2 c_0+2 c_1+4d \ ,   \nonumber \\
 b_1 &=& 4 c_0-10 c_1+12 c_2 - 4d \ , \label{eqforbeta} \\
 b_{i+1} &=& 4 c_i -2(4i+5) c_{i+1}+2(i+2)(2i+3) c_{i+2} \nonumber \ ,
\end{eqnarray}
where $i \ge 1 $. Considering the asymptotic limit of $r \rightarrow \infty$ in \rref{otheransatz}, the scattering length can be determined as
\begin{eqnarray}
u_{1D}(r) \approx -d \sqrt{\pi}V_0 \left( r - \underbrace{\frac{1+V_0 c_0}{d \sqrt{\pi}V_0}}_{a_s^{1D}}  \right) \ .
\label{asdef}
\end{eqnarray}
As can be seen in  \rref{asdef}, $a_s^{1D}$ depends on only two parameters: $ c_0$ and $d$. However, 
these parameters are determined through an infinitely large system of linear equations \rrefsb{eqforbeta}. 
The $s$-wave scattering length 
can be further separated into two terms as
\begin{eqnarray}
a_s^{1D}= \frac{2}{\sqrt{\pi}V_0}+\underbrace{\frac{1+V_0  c_0-2d}{d \sqrt{\pi}V_0}}_{a_{sc}^{1D}} \ . \label{as1D}
\end{eqnarray}

Considering only the first few terms of the summation in \rref{1dschinfo}, the explicit values of the parameters $ c_k$ and 
$d$ can be obtained assuming the following expressions:
\begin{eqnarray}
 c_0 &=& \frac{1}{2}+\sum_{k=1}^{\infty} \frac{k!}{2}  b_k \ , \label{betaexpr} \\
d  &=& \frac{1}{2}+\sum_{k=1}^{\infty} \frac{(2k-1)!!}{2^{k+1}}  b_k \ . \label{gammaexpr}
\end{eqnarray}
These statements can be proofed by induction. First, we suppose that \rrefsa{betaexpr} and \rrefsb{gammaexpr} are true up to the first $n$th 
terms as
\begin{eqnarray}
 c_0^{(n)} &=& \frac{1}{2} + \sum_{k=1}^n \frac{k!}{2}  b_k \ , \label{betanexpr} \\
d^{(n)} &=& \frac{1}{2} + \sum_{k=1}^n \frac{(2k-1)!!}{2^{k+1}}  b_k \ , \label{gammanexpr}
\end{eqnarray}
where parameters $ c_0^{(n)}$ and $d^{(n)}$ gives back the original parameters $ c_0$ and $d$  as $n$ goes
to infinity. Considering a finite number of $ b_k$, \rref{eqforbeta} terminates with the following last two equations:
\begin{eqnarray}
 b_{n-1} &=& 4 c_{n-2}^{(n)}-2(4n-3) c_{n-1}^{(n)} \ , \label{betanterm} \\
 b_n  &=&  4 c_{n-1}^{(n)} \label{betantermn} \ .
\end{eqnarray}
Increasing $n$ to $n+1$$b _k$, \rrefsa{betanterm} and \rrefsb{betantermn} are supplemented with additional terms as
\begin{eqnarray}
 b_{n-1} &=& 4 c_{n-2}^{(n+1)}-2(4n-3) c_{n-1}^{(n+1)}+  \label{betanp1termnm1}  \\
&& \hspace{2.5cm} +2(2n-1)n c_n^{(n+1)} \ , \nonumber \\
 b_n &=& 4 c_{n-1}^{(n+1)}-2(4n+1) c_n^{(n+1)}  \label{betanp1termn} \ , \\
 b_{n+1} &=& 4 c_n^{(n+1)}  \label{betanp1termnp1} \ .
\end{eqnarray}
Let us express $c_n^{(n+1)}$ in \rref{betanp1termnp1} and substitute back to \rrefsa{betanp1termnm1} and \rrefsb{betanp1termn}.
If we introduce the following notations: 
\begin{align}
 b_{n-1}' &=  b_{n-1} - \frac{(2n-1)n}{2}  b_{n+1} \ , \label{alphap1} \\
 b_n' &=  b_n +\frac{4n+1}{2} b_{n+1}  \ , \label{alphap2}
\end{align}
then \rrefsa{betanp1termnm1} and \rrefsb{betanp1termn} can be expressed in the following form:
\begin{eqnarray}
b_{n-1}' &=& 4 c_{n-2}^{(n+1)}-2(4n-3) c_{n-1}^{(n+1)} \ , \label{bprimenm1} \\
 b_n' &=& 4 c_{n-1}^{(n+1)}  \label{bprimen} \ .
\end{eqnarray}
Therefore, by recognizing the similarity between the expressions \rrefsb{bprimenm1} and \rrefsb{bprimen}, and 
\rrefsb{betanterm} and \rrefsb{betantermn}, the equations  \rrefsb{betaexpr} and \rrefsb{gammaexpr} for $ c_0^{(n+1)}$ and $d^{(n+1)}$ can be extended for the $n+1$ case as
\begin{eqnarray}
 c_0^{(n+1)} &=& \frac{1}{2} + \sum_{k=1}^{n-2} \frac{k!}{2}  b_k + \label{beta0np1} \\ 
&& + \frac{(n-1)!}{2}  b_{n-1}' +
\frac{n!}{2}  b_n' \nonumber \ , \\
d^{(n+1)} &=& \frac{1}{2} + \sum_{k=1}^{n-2} \frac{(2k-1)!!}{2^{k+1}}  b_k +  \label{gammanp1} \\ 
&& + \frac{(2n-3)!!}{2^n}  b_{n-1}' +
\frac{(2n-1)!!}{2^{n+1}}  b_n' \ . \nonumber
\end{eqnarray}
Substituting back \rrefsa{alphap1} and \rrefsb{alphap2} into \rrefsa{beta0np1} and \rrefsb{gammanp1}, we obtain 
back  \rrefsa{betanexpr} and \rrefsb{gammanexpr}, but the sum goes until $n+1$ justifying \rrefsa{betaexpr} and 
\rrefsb{gammaexpr}.

Therefore, using \rrefsa{betaexpr} and \rrefsb{gammaexpr}, the correction for the scattering length \rrefsb{as1D} can be explicitly given in the following form:
\begin{eqnarray*}
a_{sc}^{1D} &=& \frac{1+\sum\limits_{k=1}^{\infty} \left(k!  b_k - \frac{(2k-1)!!}{2^{k-1}} \frac{ b_k}{V_0}\right)}
{\sqrt{\pi}\left( 1+\sum\limits_{l=1}^{\infty} \frac{(2l-1)!!}{2^l}  b_l\right)} \ .
\end{eqnarray*}
Considering the limit $V_0 \rightarrow 0$, the following identities can be derived from \rref{alphaeq}:
\begin{eqnarray*}
\lim_{V_0 \rightarrow 0}  b_k &=& 0 \ ,  \\
\lim_{V_0 \rightarrow 0} \frac{ b_k}{V_0} &=& \frac{(-1)^k}{2(2k-1)k!} \ .
\end{eqnarray*}
Using the expression above the $a_{sc}^{1D}$ can be expressed in the following simple form:
\begin{eqnarray*}
\lim_{V_0 \rightarrow 0} a_{sc}^{1D} &=& \frac{1}{\sqrt{\pi}} \left( 1 + \sum_{k=1}^{\infty} (-1)^{k+1} \frac{(2k-3)!!}{(2k)!!}\right)=\sqrt{\frac{2}{\pi}} \ ,
\end{eqnarray*}
where in the last equation we recognize the Taylor series of $\sqrt{1+x}$ at $x = 1$.


\bibliography{scattering,Fermi_gases,Bose_gas,renormalization,2Drealization,programs,book,Bertsch_parameter,footnote,2Dscat}

\begin{thebibliography}{41}
\bibitem [{uto()}]{utoPhi}%
  \BibitemOpen
  \href@noop {} {}\bibinfo {note} {The radial Schr\"odinger equation
  \rrefsb{relsch}, and the asymptotic $s$-wave scattering length expressions
  \rrefsb{assPhy}-\rrefsb{as3dnum}, are written in different forms in Ref.
  [20]. Using the relation $u(r)= r^{(n-1)/2}\Phi_{nD}(r)$, these expressions
  can be easily derived from each other.}\BibitemShut {Stop}%
\end{thebibliography}

\end{document}